\documentclass{article}
\usepackage{amsmath,amssymb,amsthm}
\numberwithin{equation}{section}

\theoremstyle{definition}

\theoremstyle{remark}

\newcommand{\beq}{\begin{eqnarray}}
\newcommand{\eeq}{\end{eqnarray}}
\newcommand{\beqnn}{\begin{eqnarray*}}
\newcommand{\eeqnn}{\end{eqnarray*}}
\newcommand{\rd}{\partial}

\newcommand{\ZZ}{\mathbf{Z}}
\newcommand{\bsa}{\boldsymbol{a}}
\newcommand{\bst}{\boldsymbol{t}}
\newcommand{\bszero}{\boldsymbol{0}}
\newcommand{\calB}{\mathcal{B}}
\newcommand{\calL}{\mathcal{L}}

\begin{document}

\title{Differential Fay identities and auxiliary linear problem 
of integrable hierarchies}
\author{Kanehisa Takasaki
\thanks{The author is grateful to Takashi Takebe 
for collaboration. This research was partially supported 
by Grant-in-Aid for Scientific Research 
No. 16340040, No. 18340061 and No. 19540179 from 
the Japan Society for the Promotion of Science.}\\
Graduate School of Human and Environmental Studies\\
Kyoto University\\
Yoshida, Sakyo, Kyoto 606-8501, Japan\\
E-mail: takasaki@math.h.kyoto-u.ac.jp}
\date{}

\maketitle

\begin{abstract}
We review the notion of differential Fay identities 
and demonstrate, through case studies, its new role in 
integrable hierarchies of the KP type.  These identities 
are known to be a convenient tool for deriving dispersionless 
Hirota equations.  We show that differential (or, in the case 
of the Toda hierarchy, difference) Fay identities play 
a more fundamental role.  Namely, they are nothing but 
a generating functional expression of the full set of 
auxiliary linear equations, hence substantially equivalent 
to the integrable hierarchies themselves.  These results 
are illustrated for the KP, Toda, BKP and DKP hierarchies. 
As a byproduct, we point out some new features of 
the DKP hierarchy and its dispersionless limit. 
\end{abstract}

\newpage

\section{Introduction}

The simplest, but most important example 
of dispersionless limit can be seen 
in the KdV equation 
\beqnn
  \rd_tu + 6u\rd_xu + \epsilon^2\rd_3^3u = 0 
\eeqnn
with a small dispersion parameter $\epsilon^2$ 
in front of the dispersion term.  
By setting $\epsilon = 0$, the third order PDE 
reduces to the first order PDE 
(the dispersionless KdV equation) 
\beqnn
  \rd_tu + 6u\rd_xu = 0. 
\eeqnn
This naive procedure changes the properties 
of solutions drastically. For example, 
since soliton solutions can exist 
in the balance of nonlinearity and dispersion, 
the dispersionless KdV equation 
no longer has soliton-like solutions. 
Nonlinearity without dispersion leads 
to the so called gradient catastrophe, 
namely, a regular initial value develops 
singularity (``shock'') with 
$|\rd_xu| = \infty$.  This means that 
the appoximation by the dispersionless 
equation breaks down at that stage.  
In the presence of a small but nonzero 
dispersion term, this singularity is 
regularized to form an oscillatory region 
called ``dispersive shock'' 
(see the review by Lax, Levermore 
and Venakides \cite{LLV-review}). 
Nevertheless the dispersionless KdV equation 
is a good approximation until 
the gradient catastrophe takes place. 
Moreover, this equation inherits many aspects 
of integrability of the KdV equation such as 
the existence of a Lax formalism, 
an infinite number of commuting flows, etc. 
We can consider a dispersionless limit 
of the KP equation 
\beqnn
  \rd_y^2u = \rd_x(\rd_tu + u\rd_xu + \epsilon^2\rd_3^3u)
\eeqnn
in a similar manner. 

Another significant example is 
the two-dimensional Toda equation.  
This is originally a system of PDEs 
\beqnn
  \rd_x\rd_y\phi_s + e^{\phi_{s+1}-\phi_s} 
  - e^{\phi_s-\phi_{s-1}} = 0 
\eeqnn
for an infinite number of fields $\phi_s 
= \phi_s(x,y)$ indexed by an integer $s \in \ZZ$.  
$x$ and $y$ are light cone coordinates of 
the two-dimensional Minkowski spacetime.  
If $s$ is a coordinate on a one-dimensional lattice, 
and $\phi_s$ is interpreted as a field 
$\phi(s) = \phi(s,x,y)$ on 
a partially discretized three-dimensional 
spacetime, we can consider a continuum limit 
as the lattice spacing $\epsilon$ tends to $0$.  
To this end, we restart from the rescaled 
Toda equation 
\beqnn
  \epsilon\rd_x\rd_y\phi(s) 
  + e^{(\phi(s+\epsilon)-\phi(s))/\epsilon} 
  - e^{(\phi(s)-\phi(s-\epsilon))/\epsilon} 
  = 0 
\eeqnn
and let $\epsilon \to 0$. This yields 
the dispersionless Toda equation
\beqnn
  \rd_x\rd_y\phi + \rd_s(e^{\rd_s\phi}) = 0. 
\eeqnn
Remarkably, this equation is also known 
in general relativity as the Boyer-Finley equation, 
which was discovered in a classification 
of selfdual spacetimes 
with a Killing symmetry \cite{BF82,GD84}.  
This equation, too, is a kind of integrable system.  

These rather naive constructions of 
dispersionless limit can be reformulated 
in a more systematic way in the framework 
of the KdV, KP and Toda hierarchies.  
The outcome are the the dispersionless 
KdV, KP and Toda hierarchies.  
These dispersionless integrable hierarchies 
share many properties with 
the original dispersive systems.  
We refer details to the review 
by Takasaki and Takebe \cite{TT-review} 
and recall a few essential features. 

Firstly, the dispersionless limit turns out 
to be an analogue of ``quasi-classical limit'' 
in quantum mechanics.  The small parameter 
$\epsilon$ amounts to the Planck constant $\hbar$. 
This analogy (quantum-classical correspondence) 
becomes particularly fruitful when one considers 
the Lax formalism in the dispersionless limit. 
In the dispersionless KP and Toda hierarchies, 
commutators of (pseudo)differential operators 
in the Lax and Zakharov-Shabat equations 
are replaced by Poisson brackets, and 
the role of auxiliary linear equations 
are played by Hamilton-Jacobi equations.  
Actually, these ideas are rather old and 
can be found in studies of the Benney hierarchy 
in the early eighties \cite{KM77,LM79,Zakharov81}. 

Secondly, one can formulate 
the quasi-classical limit in terms of 
the tau function as well.  This is based 
on ideas borrowed from random matrices, 
topological field theory and string theory 
\cite{DVV91,Dubrovin92,Krichever94}.  
Although the Hirota equations themselves 
do not survive the dispersionless limit, 
one can find an alternative framework, 
namely, ``dispersionless Hirota equations''
\cite{NKNT94,TT-review,CK95}. 
Actually, we can see a prototype of 
dispersionless Hirota equations 
in the aforementioned dispersionless 
KdV and Toda equations.  In the case of 
the dispersionless KdV equation, 
we can convert the equation to 
\beqnn
  \rd_t\rd_xF + 3(\rd_x^2F)^2 = 0 
\eeqnn
by changing variables as $u = 2\rd_x^2F$ 
and integrating the equation once 
with respect to $x$.  In the case of 
the dispersionless Toda equation, 
we can similarly derive the equation 
\beqnn
  \rd_x\rd_yF + e^{\rd_s^2F} = 0 
\eeqnn
by subtstituting $\phi = \rd_sF$.  
The new dependent variable $F$ 
may be thought of as a counterpart of 
the tau function in the dispersionless systems.  
Note that, unlike Hirota equations 
in the dispersive case, the equations 
for the $F$ function are no longer bilinear.  

Dispersionless Hirota equations have played 
a central role in recent studies on 
dispersionless integrable hierarchies.  
They were applied to various problems 
of mathematical physics such as 
interface dynamics \cite{MWWZ00}, 
associativity equations \cite{BMRWZ01} and 
string field theory \cite{BR-SFT,BKR-SFT,BS-SFT,BSSST-SFT}. 
Alongside these applications, 
mathematical aspects of dispersionless 
Hirota equations were also studied in detail 
\cite{WZ00,KKMWWZ01,Zabrodin01,Takhtajan01,MWZ02,
BKMA03,Teo03,TTZ06}.  
It has been established through these studies 
that dispersionless Hirota equations can be 
a fundamental language for dispersionless 
integrable hierarchies.  

By the way, the notion of dispersionless 
Hirota equations was first discovered 
(in the case of the KP hierarchy 
\cite[Appendix B]{TT-review}) as 
dispersionless limit of the so called 
``differential Fay identity''.  
In other words, this identity (introduced 
by Adler and van Moerbeke \cite{AvM92} 
in a quite different context) 
is a dispersive counterpart of dispersionless 
Hirota equations.  This raises a natural question: 
Can the notion of differential Fay identity 
play the same fundamental role as that of 
dispersionless Hirota equations?  

This paper presents several case studies 
on this question.  In the case of the KP hierarchy, 
this question was already answered affirmatively 
when the notion of dispersionless Hirota equations 
was first proposed \cite{TT-review}.  
It was proven therein that the differential 
Fay identity, which is a consequence of 
the KP hierarchy, is actually equivalent 
to the KP hierarchy itself.  We shall review 
this result, and point out that the most essential 
part of this result is the fact that 
the differential Fay identity is 
a generating functional expression of 
an infinite number of auxiliary linear equations. 
We shall show that the same point of view is valid 
for some other cases as well. 

This paper is organized as follows.  
Sections 2 and 3 are devoted to the most 
fundamental cases, namely, the KP and Toda hierarchies.  
All results presented here are well known and 
can be found in the literature.  The case of 
the KP hierarchy is reviewed in detail (Section 2) 
as a prototype of the subsequent cases.  
In the case of the Toda hierarchy (Section 3), 
difference analogues of differential Fay identities 
show up.  Sections 4 and 5  deal with the BKP and DKP 
hierarchies, which are relatives of the KP hierarchy.  
Most results on the BKP hierarchy (Section 4) 
are already published except for the relation 
between the differential Fay identities and 
the auxiliary linear problem.  
The results on the DKP hierarchy (Section 5) 
are mostly new.  In Section 6 we show our conclusion 
along with several remarks.

\section{KP hierarchy and differential Fay identity}

\subsection{Bilinear equations for tau function}

Let $\bst = (t_1,t_2,\ldots)$ denote the set of 
time variables of the KP hierarchy.  The first one 
$t_1$ is identified with the spatial variable $x$ 
in the Lax formalism.  

The tau function $\tau = \tau(\bst)$ of 
the KP hierarchy \cite{Sato81,SS82,DJKM-KP,SW85} 
satisfies an infinite number of 
bilinear equations.  These equations can be 
written in a compact form as 
\beq
  (D_1^4 - 4D_1D_3 + 3D_2^2)\tau\cdot\tau = 0, 
  \quad \ldots. 
\label{KP-Hirota}
\eeq
with Hirota's notation 
\beqnn
  P(D_1,D_2,\ldots)\tau\cdot\tau 
  = \left.
    P(\rd'_1 - \rd_1,\, \rd'_2 - \rd_2,\,\ldots) 
    \tau(t'_1,t'_2,\ldots)\tau(t_1,t_2,\ldots) 
    \right|_{\bst'=\bst}, 
\eeqnn
where $\rd_n$ and $\rd'_n$ denote the 
derivatives $\rd_n = \rd/\rd t_n$, 
$\rd'_n = \rd/\rd t'_n$ in the component 
of $\bst$ and $\bst' = (t'_1,t'_2,\ldots)$.  

Date, Jimbo, Kashiwara and Miwa discovered 
that these Hirota equations can be encoded to 
a single equation of the form 
\beq
  \oint\frac{dz}{2\pi i}  e^{\xi(\bst'-\bst,z)}
  \tau(\bst' - [z^{-1}])\tau(\bst + [z^{-1}]) 
  = 0, 
\label{KP-bilin-tau}
\eeq
where $\xi(\bst,z)$ and $[\alpha]$ are 
the standard notions 
\beqnn
  \xi(\bst,z) = \sum_{n=1}^\infty t_nz^n, \quad 
  [\alpha] = \left(\alpha,\frac{\alpha^2}{2},
    \ldots,\frac{\alpha^n}{n},\ldots\right), 
\eeqnn
and $\oint$ is the contour integral along 
a sufficiently large circle $|z| = R$ 
(or a formal algebraic operator extracting 
the coefficient of $z^{-1}$ of a Laurent series).  
(\ref{KP-bilin-tau}) is understood to hold 
for arbitrary values of $\bst'$ and $\bst$. 
More precisely, (\ref{KP-bilin-tau}) is 
a generating functional expression of 
an infinite number of equations that 
can be obtained by Taylor expansion 
at $\bst' = \bst$.  

One can convert (\ref{KP-bilin-tau}) 
into a form that is more directly 
related to Hirota equations.  
This is achieved by introducing 
a new set of variables 
$\bsa = (a_1,a_2,\ldots)$ and 
substituting 
\beqnn
 \bst' \to \bst - \bsa, \quad 
 \bst \to \bst + \bsa. 
\eeqnn
(\ref{KP-bilin-tau}) thereby takes the form 
\beqnn
  \oint\frac{dz}{2\pi i}e^{-2\xi(\bsa,z)} 
  \tau(\bst + \bsa + [z^{-1}]) 
  \tau(\bst - \bsa - [z^{-1}]) 
  = 0. 
\eeqnn
It is convenient to introduce 
the elementary Schur functions 
$h_n(\bst)$, $n = 0,1,\ldots$ here. 
They are  defined by an exponential 
generating function as 
\beqnn
  e^{\xi(\bst,z)} = \sum_{n=0}^\infty h_n(\bst)z^n. 
\eeqnn
One can thereby express the exponential factor 
$e^{-2\xi(\bsa,z)}$ as 
\beqnn
  e^{-2\xi(\bsa,z)} = \sum_{n=0}^\infty h_n(-2\bsa)z^n. 
\eeqnn
Moreover, the product of two shifted tau functions 
can be expressed as 
\beqnn
\begin{aligned}
  \tau(\bst + \bsa + [z^{-1}])\tau(\bst - \bsa - [z^{-1}]) 
 &= \exp\left(\sum_{n=1}^\infty 
    \Bigl(a_n + \frac{z^{-n}}{n}\Bigr)D_n
    \right)\tau(\bst)\cdot\tau(\bst) \\
 &= \sum_{n=0}^\infty 
    h_n(\tilde{D}_{\bst})z^{-n} 
    e^{\langle\bsa,D_{\bst}\rangle}
    \tau(\bst)\cdot\tau(\bst),   
\end{aligned}
\eeqnn
where 
\beqnn
  \tilde{D}_{\bst} 
  = \Bigl(D_1,\frac{D_2}{2},\ldots,\frac{D_n}{n},\ldots\Bigr), 
  \quad 
  \langle\bsa,D_{\bst}\rangle = \sum_{n=1}^\infty a_nD_n. 
\eeqnn
(\ref{KP-bilin-tau}) can be thus eventually 
converted to the Hirota form 
\beq
  \sum_{n=0}^\infty h_n(-2\bsa)h_{n+1}(\tilde{D}_{\bst})
    e^{\langle\bsa,D_{\bst}\rangle}\tau(\bst)\cdot\tau(\bst) 
  = 0. 
\label{KP-Hirota-gen}
\eeq
Note that this procedure is reversible; 
one can trace it back and recover 
(\ref{KP-bilin-tau}) from (\ref{KP-Hirota-gen}). 

The last equation (\ref{KP-Hirota-gen}) 
is a generating functional form of 
an infinite number of Hirota equations, 
which are obtained by expanding 
(\ref{KP-Hirota-gen}) in powers of $\bsa$. 
For example, the terms linear in $a_n$'s gives 
the special Hirota equations 
\beq
  \Bigl(D_1D_n - 2h_{n+1}(\tilde{D}_{\bst})\ldots\Bigr)
  \tau(\bst)\cdot\tau(\bst) = 0, \quad 
  n = 3,4,\ldots.  
\label{KP-Hirota-special}
\eeq
Note that those for $n = 1$ and $n = 2$ are 
trivial identities.  The equation for $n = 3$ 
is exactly the lowest Hirota equation in 
(\ref{KP-Hirota}).

\subsection{Fay-type identities}

The following identity, referred to as 
the Fay identity in the following, holds 
for all tau functions of the KP hierarchy: 
\begin{multline}
  (\lambda_1 - \lambda_2)(\lambda_3 - \lambda_4) 
   \tau(\bst+[\lambda_1^{-1}]+[\lambda_2^{-1}])
   \tau(\bst+[\lambda_3^{-1}]+[\lambda_4^{-1}]) \\
- (\lambda_1 - \lambda_3)(\lambda_2 - \lambda_4) 
   \tau(\bst+[\lambda_1^{-1}]+[\lambda_3^{-1}]) 
   \tau(\bst+[\lambda_2^{-1}]+[\lambda_4^{-1}]) \\
+ (\lambda_1 - \lambda_4)(\lambda_2 - \lambda_3) 
   \tau(\bst+[\lambda_1^{-1}]+[\lambda_4^{-1}]) 
   \tau(\bst+[\lambda_2^{-1}]+[\lambda_3^{-1}]) 
 = 0. 
\label{KP-Fay}
\end{multline}
$\lambda_1,\lambda_2,\lambda_3$ and $\lambda_4$ 
are arbitrary parameters.  This is a generalization 
of Fay's trisecant identities for the Riemann theta 
function and the prime form on an arbitrary 
Riemann surface.  Sato and Sato \cite{SS82} 
derived this identity and its generalizations 
as a consequence of Pl\"ucker relations 
on an underlying Grassmann manifold.  
One can derive it from the bilinear identity 
(\ref{KP-bilin-tau}) as well, thought we omit details.  
We shall show a similar procedure for 
the differential Fay identity below.  

Adler and van Moerbeke \cite{AvM92} 
obtained the differential Fay identity 
\begin{multline}
    (\lambda - \mu)\tau(\bst+[\lambda^{-1}]+[\mu^{-1}])\tau(\bst) 
  - (\lambda - \mu)\tau(\bst+[\lambda^{-1}])\tau(\bst+[\mu^{-1}]) \\
  + (\rd_1\tau)(\bst+[\lambda^{-1}])\tau(\bst+[\mu^{-1}]) 
  - (\rd_1\tau)(\bst+[\mu^{-1}])\tau(\bst+[\lambda^{-1}]) 
  = 0 
\label{KP-diffFay}
\end{multline}
by a confluence procedure letting two of $\lambda$'s 
in the Fay identity tend to $0$.  We now derive 
this identity from the bilinear identity 
(\ref{KP-bilin-tau}).  Firstly, let us 
differentiate the bilinear equation by $t'_1$.  
This gives the equation 
\beqnn
  \oint\frac{dz}{2\pi i}e^{\xi(\bst'-\bst,z)} 
  \Bigl(z\tau(\bst'-[z^{-1}]) 
  + (\rd_1\tau)(\bst'-[z^{-1}])
  \Bigr)\tau(\bst+[z^{-1}])  = 0. 
\eeqnn
Secondly, assuming that $\lambda$ and 
$\mu$ sit on the far side of the contour 
$|z| = R$ of the integral (i.e., 
$|\lambda| > R$ and $|\mu| > R$), 
we specialize $\bst'$ as 
\beqnn
  \bst' = \bst + [\lambda^{-1}] + [\mu^{-1}]. 
\eeqnn
Then, by virtue of the identity 
\beqnn
  \sum_{n=1}^\infty 
  \frac{1}{n}\Bigl(\frac{z}{w}\Bigr)^n 
  = - \log\left(1 - \frac{z}{w}\right), 
\eeqnn
the exponential factor $e^{\xi(\bst'-\bst,z)}$ 
reduces to a rational function of $z,\lambda,\mu$ as 
\beqnn
  e^{\xi(\bst'-\bst,z)} 
  = \Bigl(1 - \frac{z}{\lambda}\Bigr)^{-1} 
    \Bigl(1 - \frac{z}{\mu}\Bigr)^{-1} 
  = \frac{\lambda\mu}{(z - \lambda)(z - \mu)}. 
\eeqnn
The foregoing equation thereby becomes 
\begin{multline*}
  \oint\frac{dz}{2\pi i}
  \frac{\lambda\mu}{(z - \lambda)(z - \mu)} 
  \Bigl(z\tau(\bst+[\lambda^{-1}]+[\mu^{-1}]-[z^{-1}])\\
  + (\rd_1\tau)(\bst+[\lambda^{-1}]+[\mu^{-1}]-[z^{-1}]
  \Bigr)\tau(\bst+[z^{-1}]) 
  = 0. 
\end{multline*}
Now assuming that $\tau(\bst)$ is a holomorphic 
function of $\bst$ in a neighborhood of $\bst = \bszero$, 
we conclude that the integrand is a meromorphic 
function of $z$ in $|z| \ge R$ with poles 
at $z = \infty,\lambda,\mu$.  Therefore 
we can use residue calculus to calculate 
the contour integral as a sum of the residues 
at the poles.  This yields (\ref{KP-diffFay}).  

(\ref{KP-diffFay}) is not very convenient 
for some purposes.  It is more suggestive 
to rewrite it as 
\beq
  \frac{\tau(\bst+[\lambda^{-1}]+[\mu^{-1}])\tau(\bst)}
       {\tau(\bst+[\lambda^{-1}])\tau(\bst+[\mu^{-1}])} 
  = 1 
  - \frac{1}{\lambda-\mu}\rd_1\log
    \frac{\tau(\bst+[\lambda^{-1}])}{\tau(\bst+[\mu^{-1}])}. 
\label{KP-diffFay1}
\eeq
Moreover, one can use the differential operator 
\beqnn
  D(z) = \sum_{n=1}^\infty \frac{z^{-n}}{n}\rd_n 
\eeqnn
to rewrite it further as 
\beq
  \exp\Bigl((e^{D(\lambda)}-1)(e^{D(\mu)}-1)\log\tau\Bigr) 
  = 1 - \frac{\rd_1(e^{D(\lambda)}-e^{D(\mu)})\log\tau}{\lambda-\mu}. 
\label{KP-diffFay2}
\eeq

As proven by Takasaki and Takebe \cite{TT-review}, 
the differential Fay identity (\ref{KP-diffFay}) 
is equivalent to the KP hierarchy itself.  
A clue of the proof is a close relationship between 
the differential Fay identity and 
the auxiliary linear problem.

\subsection{Auxiliary linear equations for wave functions}

The auxiliary linear problem for the KP hierarchy 
\cite{Sato81,SS82,DJKM-KP,SW85} 
consists of the linear equations 
\beq
  (\rd_n - B_n)\Psi = 0, 
\label{KP-auxlin}
\eeq
where $B_n$'s are differential operators of the form 
\beqnn
  B_n = \rd_1^n + b_{n,2}\rd_1^{n-2} + \cdots + b_{n,n}. 
\eeqnn
These operators (the Zakharov-Shabat operators) 
satisfy the Zakharov-Shabat equations 
\beq
  [\rd_m - B_m,\, \rd_n - B_n] = 0, 
\label{KP-ZS}
\eeq
equivalently, 
\beqnn
  \rd_n(B_n) - \rd_m(B_n) + [B_m,B_n] = 0, 
\eeqnn
where $\rd_n(B_m)$ means differentiating 
the coefficients of $B_m$ by $t_n$.  
Moreover, there is a pseudo-differential operator 
(Lax operator) of the form 
\beqnn
  L = \rd_1 + \sum_{j=1}^\infty u_{j+1}\rd_1^{-j} 
\eeqnn
such that the Zakharov-Shabat operators 
are expressed as 
\beq
  B_n = (L^n)_{\ge 0}, 
\label{KP-B-L}
\eeq
where $(\quad)_{\ge 0}$ stands for the part 
of nonnegative powers of $\rd_1$.  
The Lax operator satisfies the Lax equations 
\beq
  \rd_n(L) = [B_n,L]. 
\label{KP-Lax}
\eeq

In analogy with quantum mechanics, 
solutions of the auxiliary linear 
equations are called ``wave functions''. 
The most fundamental wave function 
can be obtained from the tau function 
by the formula 
\beqnn
  \Psi(\bst,z) 
  = \frac{\tau(\bst - [z^{-1}])}{\tau(\bst)}
    e^{\xi(\bst,z)} 
  = \frac{e^{-D(z)}\tau(\bst)}{\tau(\bst)}
    e^{\xi(\bst,z)}. 
\eeqnn
As an immediate consequence of 
the bilinear equation (\ref{KP-bilin-tau}) 
for the tau function, this wave function 
and its dual 
\beqnn
  \Psi^*(\bst,z) 
  = \frac{\tau(\bst + [z^{-1}])}{\tau(\bst)} 
    e^{-\xi(\bst,z)} 
  = \frac{e^{D(z)}\tau(\bst)}{\tau(\bst)} 
    e^{-\xi(\bst,z)} 
\eeqnn
satisfy the bilinear equation 
\beq
  \oint\frac{dz}{2\pi i}\Psi(\bst',z)\Psi^*(\bst,z) = 0. 
\label{KP-bilin-Psi}
\eeq

We can derive the auxiliary linear equations 
(\ref{KP-auxlin}) from this bilinear equation 
as follows \cite{DJKM-review,Dickey-book}.  
Differentiating (\ref{KP-bilin-Psi}) 
by $t'_n$ yield the new bilinear equation 
\beqnn
  \oint\frac{dz}{2\pi i}
  (\rd_n\Psi)(\bst',z)\Psi^*(\bst,z) 
  = 0. 
\eeqnn
In a similar manner, bilinear equations 
of the form 
\beqnn
  \oint\frac{dz}{2\pi i}
  (B\Psi)(\bst',z)\Psi^*(\bst,z) 
  = 0 
\eeqnn
hold for for any higher order 
differential operator $B = B(\rd_1)$ 
with respect to $t_1$.  (Actually, 
the bilinear equation (\ref{KP-bilin-Psi}) 
should be interpreted as 
a generating functional form of 
all those equations.)   Consequently, 
we obtain the equation 
\beq
  \oint\frac{dz}{2\pi i}R(\bst,z)\Psi^*(\bst,z) = 0 
\label{KP-remainder1}
\eeq
for the linear combination 
\beqnn
  R(\bst,z) = \rd_n\Psi(\bst,z) - B\Psi(\bst,z). 
\eeqnn
We now choose $B$ to be a differential operator 
$B_n = B_n(\rd_1)$ of the aforementioned form 
such that 
\beq
  R(\bst,z) 
  = \rd_n\Psi(\bst,z) - B_n(\rd_1)\Psi(\bst,z) 
  = O(z^{-1})e^{\xi(\bst,z)}, 
\label{KP-remainder2}
\eeq
namely, the Laurent expansion of 
$R(\bst,z)e^{-\xi(\bst,z)}$ at $z = \infty$ 
is of order $O(z^{-1})$.  Such an operator 
$B_n(\rd_1)$ does exist and is unique 
(see below for an explicit construction 
of this operator).  On the other hand, 
there is a general formula that relates 
a contour integral of the form 
\beqnn
  \oint\frac{dz}{2\pi i}(P(x',\rd_x')e^{x'z}) 
    (Q(x,\rd_x)e^{-xz}) 
\eeqnn
to the product of $P(x,\rd_x)$ and 
$Q(x,\rd_x)^*$ (the formal adjoint of 
$Q(x,\rd_x)$ \cite{DJKM-review,Dickey-book}. 
With the aid of that formula, we can prove that 
any function $R(\bst,z)$ with these properties 
(\ref{KP-remainder1}) and (\ref{KP-remainder2}) 
has to vanish identically.  Thus $\Psi(\bst,z)$ 
turns out to satisfy the auxiliary linear 
equations (\ref{KP-auxlin}).  

Having obtained the auxiliary linear equations 
for $\Psi(\bst,z)$, we can deduce the existence 
of the Lax operator as follows. 
A clue is the so called dressing (or Sato-Wilson) 
operator 
\beqnn
  W = 1 + \sum_{j=1}^\infty w_j\rd_1^{-j} 
\eeqnn
defined by the coefficients of 
the Laurent expansion 
\beqnn
  \Psi(\bst,z)e^{-\xi(\bst,z)} 
  = 1 + \sum_{j=1}^\infty w_jz^{-j} 
\eeqnn
of $\Psi(\bst,z)e^{-\xi(\bst,z)}$ at $z = \infty$.  
Note that $\Psi(\bst,z)$, in turn, 
can be expressed as 
\beq
  \Psi(\bst,z) = We^{\xi(\bst,z)}. 
\label{KP-Psi-W}
\eeq
One can immediately see that the operators 
\beqnn
  B_n = (W\rd_1^nW^{-1})_{\ge 0}
\eeqnn
turn out to fulfill the condition 
(\ref{KP-remainder2}).  
Therefore if we define $L$ as 
\beqnn
  L = W\rd_1W^{-1}, 
\eeqnn
the foregoing expression (\ref{KP-B-L}) 
of $B_n$'s is achieved.  Moreover, 
by plugging (\ref{KP-Psi-W}) into 
the auxiliary linear equations, 
one finds that $W$ satisfies 
the the Sato equations 
\beq
  \rd_n(W) 
  = B_nW - W\rd_1^n 
  = - (W\rd_1^nW^{-1})_{<0}W, 
\label{KP-Sato}
\eeq
where $(\quad)_{<0}$ stands for the part 
of negative powers of $\rd_1$.  
The Lax equations (\ref{KP-Lax}) readily 
follow from the Sato equations.  

Note that the identification of 
the Lax and Zakharov-Shabat operators 
is independent of the foregoing derivation 
of the auxiliary linear equations.  
Namely, once one can anyhow show 
that $\Psi(\bst,z)$ satisfies 
the auxiliary linear equations (\ref{KP-auxlin}), 
one can thereby conclude the existence of 
an operator $L$ which satisfies the Lax equations 
(\ref{KP-Lax}) and by which $B_n$'s are expressed 
as (\ref{KP-B-L}).  This remark becomes 
technically important when we derive 
the auxiliary linear equations from 
the differential Fay identity.

\subsection{Differential Fay identity and auxiliary linear equations}

The foregoing derivation of the auxiliary 
linear equations (\ref{KP-auxlin}) from 
the bilinear equation (\ref{KP-bilin-Psi}) 
is somewhat indirect.  We now show 
that the differential Fay identity 
gives a more direct approach to 
the auxiliary linear equations.  

Let us anyway rewrite the differential 
Fay identity in the language of the wave function.  
The first step is to shift $\bst$ 
in (\ref{KP-diffFay1}) as 
$\bst \to \bst - [\lambda^{-1}] - [\mu^{-1}]$.  
(\ref{KP-diffFay1}) thereby changes as 
\beqnn
  \frac{\tau(\bst-[\lambda^{-1}]-[\mu^{-1}])\tau(\bst)}
       {\tau(\bst-[\lambda^{-1}])\tau(\bst-[\mu^{-1}])} 
  = 1 
  + \frac{1}{\lambda-\mu}\rd_1\log
    \frac{\tau(\bst-[\lambda^{-1}])}{\tau(\bst-[\mu^{-1}])}. 
\eeqnn
The next step is to multiply both hand sides 
of the equation by 
$(\lambda-\mu)e^{\xi(\bst,\mu)}\tau(\bst-[\mu^{-1}])/
\tau(\bst)$.  This yields the equation 
\begin{multline*}
    (\lambda-\mu)e^{\xi(\bst,\mu)}
    e^{-D(\lambda)}\frac{\tau(\bst-[\mu^{-1}])}{\tau(\bst)} \\
  = \left(\lambda 
      + \rd_1\log\frac{\tau(\bst-[\lambda^{-1}])}{\tau(\bst)} 
      - \mu 
      - \rd_1\log\frac{\tau(\bst-[\mu^{-1}])}{\tau(\bst)} 
    \right) 
    \times \\
    \times 
    e^{\xi(\bst,\mu)}\frac{\tau(\bst-[\mu^{-1}])}{\tau(\bst)}. 
\end{multline*}
Noting that the commutator of $e^{-D(\lambda)}$ 
and $e^{\xi(\bst,\mu)}$ yields the factor 
$1 - \mu/\lambda$, we can rewrite 
the last equation as 
\beq
  \lambda e^{-D(\lambda)}\Psi(\bst,\mu) 
  = (\rd_1\log\Psi(\bst,\lambda) - \rd_1)\Psi(\bst,\mu). 
\label{KP-auxlin-gen}
\eeq
Expanded in powers of $\lambda$, 
this equation yields an infinite number 
of linear equations of the form 
\beq
  h_n(-\tilde{\rd}_{\bst})\Psi(\bst,\mu) 
  = v_n\Psi(\bst,\mu) 
\label{KP-auxlin-SS}
\eeq
for $n = 2,3,\ldots$, where 
$\tilde{\rd}_{\bst}$ is defined as 
\beqnn
  \tilde{\rd}_{\bst} 
  = \Bigl(\rd_1,\frac{\rd_2}{2},\ldots,\frac{\rd_n}{n},\ldots\Bigr), 
\eeqnn
and $v_n$'s are the coefficients of 
the Laurent expansion of 
$\rd_1\log\Psi(\bst,\lambda)$, namely, 
\beqnn
  \rd_1\log\Psi(\bst,\lambda) 
  = \lambda + \sum_{n=1}^\infty v_{n+1}\lambda^{-n}. 
\eeqnn

Linear equations (\ref{KP-auxlin-SS}) were 
first discovered by Sato and Sato \cite{SS82}, 
but implications of these equations remained 
to be clarified for years.  It was then pointed out 
by Takasaki and Takebe \cite{TT-review} that 
(\ref{KP-auxlin-SS}) are equivalent to the standard 
auxiliary linear problem (\ref{KP-auxlin}).  
As one can immediately see from the first few equations 
\beqnn
\begin{gathered}
  \Bigl(-\frac{1}{2}\rd_2 + \frac{1}{2}\rd_1^2\Bigr)\Psi(\bst,\mu) 
  = v_2\Psi(\bst,\mu), \\
  \Bigl(-\frac{1}{3}\rd_3 + \frac{1}{2}\rd_2\rd_1 
    - \frac{1}{6}\rd_1^3\Bigr)\Psi(\bst,\mu) 
  = v_3\Psi(\bst,\mu), \quad \ldots,\\
\end{gathered}
\eeqnn
the left hand side of the $n$-th equation 
of (\ref{KP-auxlin-SS}) takes such a form as 
\beqnn
  h_n(-\tilde{\rd}_{\bst})\Psi(\bst,\mu) 
 = \left(-\frac{1}{n}\rd_n + C_n(\rd_1,\ldots,\rd_{n-1})
   \right)\Psi(\bst,\mu), 
\eeqnn
where $C_n(\rd_1,\cdots,\rd_{n-1})$ is a differential 
operator that contains $\rd_1,\ldots,\rd_{n-1}$. 
One can recursively eliminate the operators 
$\rd_2,\ldots,\rd_{n-1}$ other than $\rd_1$ 
using the preceding equations 
$h_m(-\tilde{\rd}_{\bst})\Psi(\bst,z) 
= v_m\Psi(\bst,z)$ for $m < n$.  
The outcome are linear equations of the usual 
evolutionary form 
\beqnn
\begin{gathered}
  \rd_2\Psi(\bst,\mu) 
  = (\rd_1^2 - 2v_2)\Psi(\bst,\mu),\\
  \rd_3\Psi(\bst,\mu) 
  = (\rd_1^3 - 3v_2\rd_1 - 2\rd_1v_2 - 3v_3)\Psi(\bst,\mu), 
  \quad \ldots. 
\end{gathered}
\eeqnn
One can identify the operators 
on the right hand side to be the standard 
Zakharov-Shabat operators (\ref{KP-B-L}) 
by the same reasoning as we have used 
for deriving the auxiliary linear equations 
(\ref{KP-auxlin}) from the bilinear equation 
(\ref{KP-bilin-Psi}).  Thus (\ref{KP-auxlin-SS}) 
turns out to be equivalent to (\ref{KP-auxlin}). 
In other words, (\ref{KP-auxlin-gen}) 
is a generating functional form of 
these linear equations.  

We are thus led to the conclusion that 
the differential Fay identity is actually 
the auxiliary linear problem in disguise.  
This fact lies in the heart of the proof 
\cite{TT-review} of the fact that 
the differential Fay identity is equivalent 
to the full system of the KP hierarchy. 

Let us note that $v_n$'s are 
rather well known quantities, namely, 
conserved densities of the KP hierarchy 
\cite{Flaschka83}.  
Since $L\Psi(\bst,\lambda) 
= \lambda\Psi(\bst,\lambda)$, 
one can rewrite the foregoing defining equation 
of $v_n$'s as 
\beqnn
\begin{aligned}
  \rd_1\Psi(\bst,\lambda) 
 &= \Bigl(\lambda + \sum_{n=1}^\infty v_{n+1}\lambda^{-n} 
    \Bigr)\Psi(\bst,\lambda) \\
 &= \Bigl(L + \sum_{n=1}^\infty v_{n+1}L^{-n}
    \Bigr)\Psi(\bst,\lambda).  
\end{aligned}
\eeqnn
This implies the operator identity 
\beq
  \rd_1 = L + \sum_{n=1}^\infty v_{n+1}L^{-n}. 
\label{KP-vn-L}
\eeq
Moreover, since 
\beqnn
  \log\Psi(\bst,\lambda) 
  = \xi(\bst,\lambda) + (e^{-D(\lambda)}-1)\log\tau, 
\eeqnn
one has the explicit formula 
\beq
  v_n = \rd_1h_n(-\tilde{\rd}_{\bst})\log\tau. 
\label{KP-vn-tau}
\eeq

The differential Fay identity is particularly 
useful for studying the limit to 
the dispersionless KP hierarchy.  
This limit is achieved by a kind of 
``quasi-classical'' limit.  One can formulate 
this procedure in both the Hirota formalism 
and the Lax formalism.  Let us now turn 
to this subject.

\subsection{Quasi-classical limit in Hirota formalism}

In the Hirota formalism, the quasi-classical limit 
is formulated as follows \cite{TT96}.  
Firstly, we allow the tau function to depend 
on an extra small parameter $\hbar$ 
(an analogue of the Planck constant 
in quantum mechanics) as $\tau = \tau(\hbar,\bst)$.  
Secondly, we assume that the rescaled tau function 
$\tau_\hbar(\bst) = \tau(\hbar,\hbar^{-1}\bst)$ 
behaves as 
\beq
  \tau_\hbar(\bst) 
  = \exp\Bigl(\hbar^{-2}F(\bst) + O(\hbar^{-1})\Bigr) 
\label{KP-tau-hbar}
\eeq
in the limit of $\hbar \to 0$.  
$F(\bst)$ is a key function in the theory 
of dispersionless integrable systems, 
having several different names such as 
``dispersionless tau function'', 
``free energy'', ``prepotential'', etc. 
that stem from its origin in 
problems of mathematical physics 
\cite{DVV91,Dubrovin92,Krichever94}. 

By rescaling the tau function as above, 
the differential Fay identity (\ref{KP-diffFay2}) 
takes the rescaled form 
\beqnn
  \exp\Bigl((e^{\hbar D(\lambda)}-1)
            (e^{\hbar D(\mu)}-1)\log\tau_\hbar\Bigr) 
  = 1 
  - \hbar\frac{\rd_1(e^{\hbar D(\lambda)}-e^{\hbar D(\mu)})
       \log\tau_\hbar}{\lambda-\mu}. 
\eeqnn
Under the foregoing ansatz (\ref{KP-tau-hbar}), 
both hand sides of this equation 
has a finite limit as $\hbar \to 0$.  
This yields the equation \cite{TT-review} 
\beq
  e^{D(\lambda)D(\mu)F} 
  = 1 - \frac{\rd_1(D(\lambda)-D(\mu))F}{\lambda-\mu} 
\label{dKP-Hirota}
\eeq
for the $F$ function $F = F(\bst)$.  

The last equation, known as the dispersionless 
Hirota equation,  is a generating functional form 
of an infinite number of equations, which are 
obtained by Laurent expansion of both hand sides 
in powers of $\lambda$ and $\mu$.  
It should be noted that ``the dispersionless 
Hirota equation'' might be a misleading name, 
because (\ref{dKP-Hirota}) is neither 
a direct limit of the Hirota equations 
(\ref{KP-Hirota}) themselves nor of 
their generating functional form 
(\ref{KP-bilin-tau}), but of the differential 
Fay identity.  Therefore, it would be better 
to call (\ref{dKP-Hirota}) ``the dispersionless 
differential Fay identity''. 

In this respect, Carroll and Kodama \cite{CK95} 
observed an interesting fact:  They pointed out 
that the special Hirota equations 
(\ref{KP-Hirota-special}) have 
a direct quasi-classical limit.  
To see this fact, it is convenient 
to start from the generating functional form 
\beqnn
  2e^{D(\mu)}\tau\cdot\tau 
  = 2 + \sum_{n=1}^\infty \mu^{-n-1}D_1D_n\tau\cdot\tau 
\eeqnn
of (\ref{KP-Hirota-special}).  It is not hard to see 
that this equation survives the quasi-classical limit 
and yields the equation 
\beq
  e^{D(\mu)^2F} 
  = 1 + \sum_{n=1}^\infty \mu^{-n-1}\rd_1\rd_nF  
\label{dKP-Hirota-special1}
\eeq
for the $F$ function.  Actually, 
the last equation can also be obtained 
by letting $\lambda \to \mu$ in (\ref{dKP-Hirota}).  
Expanded in powers of $\mu$, this equation 
generates an infinite number of equations 
of the form 
\beq
  \rd_1\rd_nF = h_{n+1}(Z_1,Z_2,\ldots), \quad 
  n = 1,2,\ldots, 
\label{dKP-Hirota-specia2}
\eeq
where 
\beqnn
  Z_n = \sum_{j+k=n}\frac{\rd_j\rd_kF}{jk}. 
\eeqnn
Carroll and Kodama studied the algebraic 
structure of these genuine 
``dispersionless Hirota equations'' in detail.  

As the differential Fay identity is equivalent 
to the KP hierarchy itself, its dispersionless analogue 
(\ref{dKP-Hirota}) turns out to be equivalent 
to (the Lax formalism of) the dispersionless 
KP hierarchy \cite{BMRWZ01,BKMA03,Teo03}.  
This is by no means an obvious fact. 
We shall return to this issue later on.

\subsection{Quasi-classical limit in Lax formalism}

In the Lax formalism, the procedure of 
quasi-classical limit starts from the WKB ansatz 
\cite{KG88-89,Zakharov94} 
\beq
  \Psi_\hbar(\bst,z) 
  = \exp\Bigl(\hbar^{-1}S(\bst,z) + O(\hbar^0)\Bigr) 
\label{KP-Psi-hbar}
\eeq
of the rescaled wave function 
$\Psi_\hbar(\bst,z) = \Psi(\hbar,\hbar^{-1}\bst,z)$.  
It is easy to see that this WKB ansatz 
follows from the quasi-classical ansatz 
(\ref{KP-tau-hbar}) of the tau function. 
The associated $S$ function is given by 
\beqnn
  S(\bst,z) = \xi(\bst,z) - D(z)F 
  = \sum_{n=1}^\infty t_nz^n 
    - \sum_{n=1}^\infty \frac{z^{-n}}{n}\rd_nF. 
\eeqnn

The auxiliary linear equations for 
the rescaled wave function takes 
the rescaled form 
\beqnn
  \hbar\rd_n\Psi_\hbar(\bst,z) 
  = B_{\hbar,n}(\hbar\rd_1)\Psi_\hbar(\bst,z), 
\eeqnn
where the coefficients of the operators 
\beqnn
  B_{\hbar,n}(\hbar\rd_1) 
  = (\hbar\rd_1)^n + b_{\hbar,n,2}(\hbar\rd_1)^{n-2} 
    + \cdots + b_{\hbar,n,n} 
\eeqnn
have a smooth limit as $\hbar \to 0$.  
These equations may be thought of as analogues 
of the time-dependent Schr\"odinger equations 
in quantum mechanics.  Consequently, 
the $S$ function satisfies 
the Hamilton-Jacobi equations 
\beq
  \rd_nS(\bst,z) = \calB_n(\rd_1S(\bst,z)) 
\label{dKP-HJ}
\eeq
with the Hamiltonians 
\beqnn
  \calB_n(p) = \lim_{\hbar\to 0}B_{\hbar,n}(p). 
\eeqnn

These stuff resemble the setup of 
quasi-classical approximation in quantum mechanics.  
The Zakharov-Shabat operators $B_{\hbar,n}
= B_{\hbar,n}(\hbar\rd_1)$ are now replaced 
by functions $\calB_n = \calB_n(p)$ of 
a new variable $p$, which may be interpreted 
as the conjugate momentum of the coordinate $x = t_1$.  
By this ``quantum-classical correspondence'', 
commutators of differential operators in 
the Lax and Zakharov-Shabat equations of 
the KP hierarchy are replaced by 
Poisson brackets 
\beqnn
  \{F,G\} 
  = (\rd_pF)(\rd_1G) - (\rd_1F)(\rd_pG) 
\eeqnn
of functions on a two-dimensional phase space. 
We are thus led to the Zakharov-Shabat equations 
\beq
  \rd_n(\calB_m) - \rd_m(\calB_n) + \{\calB_m,\calB_n\} = 0 
\label{dKP-ZS}
\eeq
and the Lax equations 
\beq
 \rd_n(\calL) = \{\calB_n,\calL\} 
\label{dKP-Lax}
\eeq
with respect to Poisson brackets 
\cite{KG88-89,Krichever91,TT92}.  
The Lax operator is also replaced by 
a function of the form 
\beqnn
  \calL  = p + \sum_{j=1}^\infty u_{j+1}p^{-j}. 
\eeqnn
$\calB_n$'s are thereby expressed as 
\beqnn
  \calB_n = (\calL^n)_{\ge 0}, 
\eeqnn
where $(\quad)_{\ge 0}$ denotes 
the polynomial part of Laurent series 
of $p$.  

The Hamilton-Jacobi equations (\ref{dKP-HJ}) 
and the Zakharov-Shabat-Lax equations 
(\ref{dKP-ZS}) -- (\ref{dKP-Lax}) can be 
transferred from one side to the other 
by the functional relation 
\beq
  p = \rd_1S(\bst,z) 
    = z - \sum_{n=1}^\infty \frac{z^{-n}}{n}\rd_1\rd_nF 
\label{dKP-p(z)}
\eeq
connecting the spectral parameter $z$ 
and the momentum $p$.  The inverse 
$p \mapsto z$ of the map $z \mapsto p$ 
is exactly the Lax function $\calL = \calL(p)$.  
One can see this fact from (\ref{KP-vn-tau}) 
and (\ref{KP-vn-L}) as follows.  
By rescaling the tau function, 
(\ref{KP-vn-tau}) becomes 
\beqnn
  v_{\hbar,n} 
  = \hbar\rd_1h_n(-\hbar\tilde{\rd}_{\bst})\log\tau_\hbar, 
\eeqnn
hence 
\beqnn
  \lim_{\hbar\to 0}v_{\hbar,n} 
  = - \frac{1}{n}\rd_1\rd_nF. 
\eeqnn
Consequently, (\ref{KP-vn-L}) yields 
\beqnn
  p = \calL(p) 
    - \sum_{n=1}^\infty \frac{\calL(p)^{-n}}{n}\rd_1\rd_nF 
\eeqnn
in the limit as $\hbar \to 0$.  
Comparing this relation with (\ref{dKP-p(z)}), 
one finds that the inverse of the map $z \mapsto 
p = \rd_1S(\bst,z)$ is given by 
$z = \calL(p)$.

\subsection{Dispersionless Hirota equation  and 
Hamilton-Jacobi equations}

As already mentioned, the dispersionless 
Hirota equation (\ref{dKP-Hirota}) is 
equivalent to the Lax formalism of 
the dispersionless KP hierarchy 
\cite{BMRWZ01,BKMA03,Teo03}.  
Main part of the proof of this fact 
is to show that (\ref{dKP-Hirota}) 
is a generating functional form of 
the Hamilton-Jacobi equations (\ref{dKP-HJ}). 
This is a dispersionless analogue of 
the relation between the differential Fay 
identity and the auxiliary linear equations.  

The notion of ``Faber polynomials'' 
in complex analysis \cite{Teo03} 
plays a central role here.  
Let us demonstrate its power 
by assuming (\ref{dKP-Hirota}) and 
deriving (\ref{dKP-HJ}) therefrom.  
We define $p(z)$ as 
\beqnn 
  p(z) = \rd_1S(\bst,z) 
\eeqnn
and rewrite (\ref{dKP-Hirota}) as 
\beqnn
  e^{D(\lambda)D(\mu)F} 
  = \frac{p(\lambda)-p(\mu)}{\lambda-\mu}. 
\eeqnn
Taking the logarithm of both hand sides yields 
\beqnn
  D(\lambda)D(\mu)F 
  = \log\frac{p(\lambda)-p(\mu)}{\lambda-\mu}. 
\eeqnn
By virtue of the identity 
\beqnn
  \log\Bigl(1 - \frac{\mu}{\lambda}\Bigr) 
  = - \sum_{n=1}^\infty \frac{1}{n}
      \Bigl(\frac{\mu}{\lambda}\Bigr)^n, 
\eeqnn
we can further rewrite the last equation as 
\beq
\begin{aligned}
  \log\frac{p(\lambda)-p(\mu)}{\lambda} 
 &= - \sum_{n=1}^\infty 
      \frac{\lambda^{-n}}{n} 
      \left(\mu^n - \sum_{m=1}^\infty\frac{\mu^{-m}}{m}\rd_n\rd_mF 
      \right) \\
 &= - \sum_{n=1}^\infty \frac{\lambda^{-n}}{n}\rd_n S(\mu). 
\end{aligned}
\label{dKP-HJ-gen}
\eeq
The proble is to solve this equation 
for $\rd_nS(\mu)$.  To this end, 
we introduce the Farber polynomials 
$\Phi_n(p)$, $n = 1,2,\ldots$, of the map 
$z \mapsto p(z)$, which are uniquely 
defined by the generating functional relation 
\beqnn
  \log\frac{p(z) - q}{z} 
  = - \sum_{n=1}^\infty 
      \frac{z^{-n}}{n}\Phi_n(q). 
\eeqnn
More explicitly, as we shall show below, 
$\Phi_n(p)$ are given by the polynomial part 
of the inverse map $p \mapsto z = z(p)$ as 
\beq
  \Phi_n(p) = \Bigl(z(p)^n\Bigr)_{\ge 0}. 
\label{dKP-Faber}
\eeq
In the present setting, $z(p)$ is 
nothing but $\calL(p)$, hence 
\beqnn
  \Phi_n(p) = \calB_n(p). 
\eeqnn
We can now solve (\ref{dKP-HJ-gen}) 
for $\rd_nS(\mu)$ as 
\beq
  \rd_nS(\mu) = \Phi_n(p(\mu)) = \calB_n(p(\mu)).  
\label{dKP-HJ-Faber}
\eeq
The last equations are exactly the Hamilton-Jacobi 
equations (\ref{dKP-HJ}). 

Let us show that (\ref{dKP-Faber}) 
indeed holds.  To this end, we differentiate 
the defining equation of $\Phi_n(q)$ by $z$. 
This yields the identity 
\beqnn
  \frac{p'(z)}{p(z) - q} 
  = z^{-1} + \sum_{n=1}^\infty z^{-n-1}\Phi_n(q), 
\eeqnn
where $p'(z)$ denotes the $z$-derivative 
$\rd_zp(z)$.  We can extract $\Phi_n(q)$ 
by a contour integral as 
\beqnn
  \Phi_n(q) 
  = \oint\frac{dz}{2\pi i} \frac{z^np'(z)}{p(z) - q}, 
\eeqnn
where the contour of integral is a sufficiently 
large circle $|z| = R$ such that $|q| < R$.   
Since the map $z \mapsto p(z)$ is invertible 
in a neighborhood of $z = \infty$, 
we choose the contour inside that neighborhood 
and change the variable from $z = z(p)$ to $p$ as 
\beqnn
  \Phi_n(q) 
  = \oint\frac{dp}{2\pi i}\frac{z(p)^n}{p - q}. 
\eeqnn
The contour of the transformed integral 
is a closed curve $C$ that encircles 
$p = \infty$ but not $q$.  This integral 
is nothing but the polynomial part of $z(q)^n$.

\section{Toda hierarchy and difference Fay identity}

\subsection{Bilinear equations for tau function} 

The Toda hierarchy \cite{UT84} 
has a discrete variable $s \in \ZZ$ 
(lattice coordinate) and two sets of continuous 
time variables $\bst = (t_1,t_2,\ldots)$, 
$\bar{\bst} = (\bar{t}_1,\bar{t}_2,\ldots)$.  
The tau function $\tau = \tau(s,\bst,\bar{\bst})$ 
satisfies an infinite number of Hirota equations 
\beq
\begin{gathered}
 D_1\bar{D}_1\tau(s,\bst,\bar{\bst})\cdot\tau(s,\bst,\bar{\bst}) 
  + 2\tau(s+1,\bst,\bar{\bst})\tau(s-1,\bst,\bar{\bst}) = 0, \\
 (D_2 + D_1^2)
  \tau(s+1,\bst,\bar{\bst})\cdot\tau(s,\bst,\bar{\bst}) = 0, \\
 (\bar{D}_2 + \bar{D}_1^2)
  \tau(s,\bst,\bar{\bst})\cdot\tau(s+1,\bst,\bar{\bst}) = 0, 
  \quad \ldots, 
\end{gathered}
\label{Toda-Hirota}
\eeq
where $D_n$ and $\bar{D}_n$ are 
the Hirota bilinear operators associated 
with $\rd_n = \rd/\rd t_n$ and 
$\bar{\rd}_n = \rd/\rd\bar{t}_n$. 
A generating functional form of these Hirota equations 
is given by the bilinear equation 
\begin{multline}
   \oint\frac{dz}{2\pi i}z^{s'-s}e^{\xi(\bst'-\bst,z)} 
      \tau(s',\bst'-[z^{-1}],\bar{\bst}') 
      \tau(s,\bst+[z^{-1}],\bar{\bst}) \\
  = \oint\frac{dz}{2\pi i}z^{s'-s}e^{\xi(\bar{\bst}'-\bar{\bst},z^{-1})} 
      \tau(s'+1,\bst',\bar{\bst}'-[z]) 
      \tau(s-1,\bst,\bar{\bst}+[z]). 
\label{Toda-bilin-tau}
\end{multline}
Whereas the contour of integral on 
the left hand side is a sufficiently large 
circle $|z| = R$, that of the right hand side 
is a sufficiently small circle $|z| = R^{-1}$ 
around the origin (or one may understand 
both integrals as a purely algebraic 
operator extracting the coefficient of $z^{-1}$).  

It is easy to from this bilinear equation 
that the Toda hierarchy contains the KP and 
modified KP hierarchies \cite{DJKM-KP} 
as subsystems.  For example, if $\bst'$ and $s$ 
are specialized as  $\bar{\bst}'=\bar{\bst}$ 
and $s' = s$,  (\ref{Toda-bilin-tau}) 
reduces to the bilinear equation 
\beqnn
  \oint\frac{dz}{2\pi i}e^{\xi(\bst'-\bst,z)} 
  \tau(s,\bst'-[z^{-1}],\bar{\bst}) 
  \tau(s,\bst+[z^{-1}],\bar{\bst}) 
  = 0, 
\eeqnn
which is substantially the bilinear equation 
for the KP hierarchy.  Thus the tau function 
of the Toda hierarchy, viewed as a function of $\bst$, 
is a tau function of the KP hierarchy as well.  
The last equation can be generalized to 
the bilinear equations 
\beq
  \oint\frac{dz}{2\pi i}e^{\xi(\bst'-\bst,z)} 
  \tau(s',\bst'-[z^{-1}],\bar{\bst}) 
  \tau(s,\bst+[z^{-1}],\bar{\bst}) 
  = 0 
  \quad \mbox{for $s' \ge s$}, 
\label{mKP-bilin-tau}
\eeq
which show that the tau function 
may be thought of as the tau function 
of the modified KP hierarchy 
with respect to $\bst$ and $s$.  
The same interpretation holds true 
for $\bar{\bst}$ and $s$.  

As in the case of the KP hierarchy, 
one can rewrite this bilinear equation 
into the Hirota form 
\begin{multline}
  \sum_{n=0}^\infty 
    h_n(-2\bsa)h_{n+s'-s+1}(\tilde{D}_{\bst})
    e^{\langle\bsa,D_{\bst}\rangle 
       + \langle\bar{\bsa},D_{\bar{\bst}}\rangle} 
    \tau(s,\bst,\bar{\bst})\cdot\tau(s',\bst,\bar{\bst}) \\
  = \sum_{n=0}^\infty 
    h_n(-2\bar{\bsa})
    h_{n-s'+s-1}(\tilde{D}_{\bar{\bst}})
    e^{\langle\bsa,D_{\bst}\rangle 
       + \langle\bar{\bsa},D_{\bar{\bst}}\rangle} 
    \tau(s-1,\bst,\bar{\bst})\cdot\tau(s'+1,\bst,\bar{\bst}), 
\label{Toda-Hirota-gen}
\end{multline}
where $\bsa = (a_1,a_2,\ldots)$ and 
$\bar{\bsa} = (\bar{a}_1,\bar{a}_2,\ldots)$ 
are new variables, and $\tilde{D}_{\bar{\bst}}$ 
and $\langle\bar{\bsa},D_{\bar{\bst}}\rangle$ are 
the counterparts of $\tilde{D}_{\bst}$ 
and $\langle\bsa,D_{\bst}\rangle$ for $\bar{\bst}$, 
namely, 
\beqnn
  \tilde{D}_{\bar{\bst}} 
  = \Bigl(\bar{D}_1,\frac{\bar{D}_2}{2},\ldots, 
          \frac{\bar{D}_n}{n},\ldots\Bigr), \quad 
  \langle\bar{\bsa},\bar{D}_{\bar{\bst}}\rangle 
  = \sum_{n=1}^\infty \bar{a}_n\bar{D}_n. 
\eeqnn 
Expanded in powers of $\bsa$ and $\bar{\bsa}$, 
(\ref{Toda-Hirota-gen}) generates an infinite number 
of Hirota equations.

\subsection{Difference Fay identities}

We can derive three Fay-type identities 
\cite{Zabrodin01,Teo06} from (\ref{Toda-bilin-tau}) 
by specializing $s',\bst',\bar{\bst}'$ as follows. 
\begin{itemize}
\item[(i)] $s' = s+1$, $\bst' = \bst+[\lambda^{-1}]+[\mu^{-1}]$, 
$\bar{\bst}' = \bar{\bst}$. 
\item[(ii)] $s' = s-3$, $\bst' = \bst$, 
$\bar{\bst}' = \bar{\bst}+[\lambda]+[\mu]$, 
\item[(iii)] $s' = s$, $\bst' = \bst+[\lambda^{-1}]$. 
$\bar{\bst}' = \bar{\bst}+[\mu]$. 
\end{itemize}
As we have shown in the case of the KP hierarchy, 
the exponential factors in the bilinear identity 
thereby become rational functions, 
and calculation of the contour integrals 
reduces to residue calculus. 

Note that the role of $x = t_1$ in the KP hierarchy 
is now played by $s$.  Unlike that case, 
however, we now shift $s$ rather than 
differentiate the bilinear equation.  
For this reason, we call the following 
Fay-type identities ``difference Fay identities''.  

\paragraph*{(i)} 
The specialized bilinear equation becomes 
\begin{multline*}
  \oint\frac{dz}{2\pi i}z\frac{\lambda\mu}{(z-\lambda)(z-\mu)}
   \tau(s+1,\bst+[\lambda^{-1}]+[\mu^{-1}]-[z^{-1}],\bar{\bst})
   \tau(s,\bst+[z^{-1}],\bar{\bst}) \\
  =\oint\frac{dz}{2\pi i}
    z\tau(s+2,\bst,\bar{\bst}-[z]) 
    \tau(s-1,\bst,\bar{\bst}+[z]). 
\end{multline*}
Residue calculus yields the first difference Fay identity 
\begin{multline}
  \tau(s+1,\bst+[\lambda^{-1}]+[\mu^{-1}])
   \tau(s,\bst,\bar{\bst}) \\
 - \frac{\lambda}{\lambda-\mu}
   \tau(s+1,\bst+[\mu^{-1}],\bar{\bst})
   \tau(s,\bst+[\lambda^{-1}],\bar{\bst}) \\
 + \frac{\mu}{\lambda-\mu} 
   \tau(s+1,\bst+[\lambda^{-1}],\bar{\bst}) 
   \tau(s,\bst+[\mu^{-1}],\bar{\bst}) 
 = 0. 
\label{Toda-diffFay(i)}
\end{multline}

\paragraph*{(ii)}
The specialized bilinear equation becomes 
\begin{multline*}
  \oint\frac{dz}{2\pi i} z^{-3}
   \tau(s-3,\bst-[z^{-1}],\bar{\bst}+[\lambda]+[\mu]) 
   \tau(s,\bst+[z^{-1}],\bar{\bst}) \\
 = \oint\frac{dz}{2\pi i} z^{-3}\frac{z^2}{(z-\lambda)(z-\mu)} 
   \tau(s-2,\bst,\bar{\bst}+[\lambda]+[\mu]-[z]) 
   \tau(s-1,\bst,\bar{\bst}+[z]). 
\end{multline*}
Residue calculus yields the second difference Fay identity 
\begin{multline}
  \frac{1}{\lambda\mu}
   \tau(s-2,\bst,\bar{\bst}+[\lambda]+[\mu]) 
   \tau(s-1,\bst,\bar{\bst}) \\
 + \frac{1}{\lambda(\lambda-\mu)}
   \tau(s-2,\bst,\bar{\bst}+[\mu])
   \tau(s-1,\bst,\bar{\bst}+[\lambda]) \\
 - \frac{1}{\mu(\lambda-\mu)}\tau(s-2,\bst,\bar{\bst}+[\lambda]) 
   \tau(s-1,\bst,\bar{\bst}+[\mu]) 
 = 0. 
\label{Toda-diffFay(ii)}
\end{multline}

\paragraph*{(iii)} 
The specialized bilinear equation becomes 
\begin{multline*}
   \oint\frac{dz}{2\pi i} \frac{-\lambda}{z-\lambda} 
   \tau(s,\bst+[\lambda^{-1}]-[z^{-1}],\bar{\bst}+[\mu]) 
   \tau(s,\bst+[z^{-1}],\bar{\bst}) \\
 = \oint\frac{dz}{2\pi i} \frac{z}{z-\mu} 
   \tau(s+1,\bst+[\lambda^{-1}],\bar{\bst}+[\mu]-[z]) 
   \tau(s-1,\bst,\bar{\bst}+[z]). 
\end{multline*}
Residue calculus yields the third difference Fay identity: 
\begin{multline}
   \lambda\tau(s,\bst+[\lambda^{-1}],\bar{\bst}+[\mu]) 
   \tau(s,\bst,\bar{\bst}) \\
 - \lambda\tau(s,\bst,\bar{\bst}+[\lambda^{-1}],\bar{\bst}) 
   \tau(s,\bst,\bar{\bst}+[\mu]) \\
 + \mu\tau(s+1,\bst+[\lambda^{-1}],\bar{\bst}) 
   \tau(s-1,\bst,\bar{\bst}+[\mu]) 
 = 0. 
\label{Toda-diffFay(iii)}
\end{multline}

These difference Fay identities can be cast 
into different forms.  The following are 
counterparts of (\ref{KP-diffFay1}): 
\begin{multline}
   \frac{\tau(s,\bst+[\lambda^{-1}]+[\mu^{-1}],\bar{\bst})
         \tau(s,\bst,\bar{\bst})}
        {\tau(s,\bst+[\lambda^{-1}],\bar{\bst}) 
         \tau(s,\bst+[\mu^{-1}],\bar{\bst})} 
 = \frac{\tau(s,\bst,\bar{\bst})}
        {(\lambda-\mu)\tau(s-1,\bst,\bar{\bst})} 
   \times \\
   \times\left(
       \frac{\lambda\tau(s-1,\bst+[\lambda^{-1}],\bar{\bst})}
            {\tau(s,\bst+[\lambda^{-1}],\bar{\bst})} 
     - \frac{\mu\tau(s-1,\bst+[\mu^{-1}],\bar{\bst})} 
            {\tau(s,\bst+[\mu^{-1}],\bar{\bst})} 
   \right), 
\label{Toda-diffFay1(i)}
\end{multline}
\begin{multline}
   \frac{\tau(s,\bst,\bar{\bst}+[\lambda]+[\mu])\tau(s,\bst,\bar{\bst})}
     {\tau(s,\bst,\bar{\bst}+[\lambda])\tau(s,\bst,\bar{\bst}+[\mu])}
 = \frac{\tau(s,\bst,\bar{\bst})}
        {(\lambda-\mu)\tau(s+1,\bst,\bar{\bst})} 
   \times \\
   \times\left( 
     - \frac{\mu\tau(s+1,\bst,\bar{\bst}+[\lambda])} 
            {\tau(s,\bst,\bar{\bst}+[\lambda])}
     + \frac{\lambda\tau(s+1.\bst,\bar{\bst}+[\mu])} 
            {\tau(s,\bst,\bar{\bst}+[\mu])} 
   \right), 
\label{Toda-diffFay1(ii)}
\end{multline}
\begin{multline}
  \frac{\tau(s,\bst+[\lambda^{-1}],\bar{\bst}+[\mu])
         \tau(s,\bst,\bar{\bst})} 
        {\tau(s,\bst+[\lambda^{-1}],\bar{\bst}) 
         \tau(s,\bst,\bar{\bst}+[\mu])} \\
  = 1 
  - \frac{\mu}{\lambda} 
    \frac{\tau(s+1,\bst+[\lambda^{-1}],\bar{\bst}) 
          \tau(s-1,\bst,\bar{\bst}+[\mu])}
         {\tau(s,\bst+[\lambda^{-1}],\bar{\bst}) 
          \tau(s,\bst,\bar{\bst}+[\mu])}. 
\label{Toda-diffFay1(iii)}
\end{multline}
Similarly, the following equations 
are counterparts of (\ref{KP-diffFay2}): 
\begin{multline}
  \exp\Bigl((e^{D(\lambda)}-1)(e^{D(\mu)}-1)\log\tau\Bigr)\\
  = \frac{1}{\lambda-\mu} 
  \Bigl(
    \lambda\exp\Bigl((e^{D(\lambda)}-1)(e^{-\rd_s}-1)\log\tau\Bigr)\\
  \mbox{}
    - \mu\exp\Bigl((e^{D(\mu)}-1)(e^{-\rd_s}-1)\log\tau\Bigr)
  \Bigr), 
\label{Toda-diffFay2(i)}
\end{multline}
\begin{multline}
  \exp\Bigl((e^{\bar{D}(\lambda)}-1)(e^{\bar{D}(\mu)}-1)\log\tau\Bigr)\\
  = \frac{1}{\lambda-\mu} 
  \Bigl(
    - \mu\exp\Bigl((e^{\bar{D}(\lambda)}-1)(e^{\rd_s}-1)\log\tau\Bigr)\\
  \mbox{} 
    + \lambda\exp\Bigl((e^{\bar{D}(\mu)}-1)(e^{\rd_s}-1)\log\tau\Bigr) 
  \Bigr), 
\label{Toda-diffFay2(ii)}
\end{multline}
\begin{multline}
  \exp\Bigl((e^{D(\lambda)}-1)(e^{\bar{D}(\mu)}-1)\log\tau\Bigr) \\ 
  = 1 
  - \frac{\mu}{\lambda}
    \exp\Bigl((e^{D(\lambda)}-1)(e^{\rd_s}-1)\log\tau 
        + (e^{\bar{D}(\mu)}-1)(e^{-\rd_s}-1)\log\tau \\
      \mbox{}
        - (e^{\rd_s}-1)(e^{-\rd_s}-1)\log\tau\Bigr), 
\label{Toda-diffFay2(iii)}
\end{multline}
where 
\beqnn
  D(z) = \sum_{n=1}^\infty \frac{z^{-n}}{n}\rd_n, \quad 
  \bar{D}(z) = \sum_{n=1}^\infty \frac{z^n}{n}\bar{\rd}_n. 
\eeqnn

\subsection{Auxiliary linear equations for wave functions}

We now introduce the wave functions 
\beqnn
\begin{gathered}
  \Psi(s,\bst,\bar{\bst},z) 
  = \frac{\tau(s,\bst-[z^{-1}],\bar{\bst})}
    {\tau(s,\bst,\bar{\bst})}
    z^se^{\xi(\bst,z)}, \\
  \bar{\Psi}(s,\bst,\bar{\bst},z) 
  = \frac{\tau(s+1,\bst,\bar{\bst}-[z])}
    {\tau(s,\bst,\bar{\bst})} 
    z^se^{\xi(\bar{\bst},z^{-1})} 
\end{gathered}
\eeqnn
and the duals 
\beqnn
\begin{gathered}
  \Psi^*(s,\bst,\bar{\bst},z) 
  = \frac{\tau(s,\bst+[z^{-1}],\bar{\bst})}
    {\tau(s,\bst,\bar{\bst})}
    z^{-s}e^{\xi(\bst,z)}, \\
  \bar{\Psi}^*(s,\bst,\bar{\bst},z) 
  = \frac{\tau(s-1,\bst,\bar{\bst}+[z])}
    {\tau(s,\bst,\bar{\bst})} 
    z^{-s}e^{\xi(\bar{\bst},z^{-1})}. 
\end{gathered}
\eeqnn
The bilinear equation (\ref{Toda-bilin-tau}) 
for the tau function becomes the bilinear 
equation 
\beq
  \oint\frac{dz}{2\pi i}
  \Psi(s',\bst',\bar{\bst}',z)
  \Psi^*(s,\bst,\bar{\bst},z) 
= \oint\frac{dz}{2\pi i}
  \bar{\Psi}(s',\bst',\bar{\bst}',z)
  \bar{\Psi}^*(s,\bst,\bar{\bst},z) 
\label{Toda-bilin-Psi}
\eeq
for these wave functions.  
From this bilinear equation, 
one can derive the auxiliary 
linear equations 
\beq
  (\rd_n - B_n)\Phi = 0, \quad 
  (\bar{\rd}_n - \bar{B}_n)\Phi = 0
\label{Toda-auxlin}
\eeq
for $\Phi = \Psi(s,\bst,\bar{\bst},z), 
\bar{\Psi}(s,\bst,\bar{\bst},z)$.  
$B_n$ and $\bar{B}_n$ are difference operators 
of the form 
\beqnn
\begin{gathered}
  B_n = e^{n\rd_s} + b_{n,1}e^{(n-1)\rd_s} 
         + \cdots + b_{n,n}, \\
  \bar{B}_n = \bar{b}_{n,0}e^{-n\rd_s} 
         + \cdots + \bar{b}_{n,n-1}e^{-\rd_s}, 
\end{gathered}
\eeqnn
and satisfy the Zakharov-Shabat equations 
\beq
\begin{gathered}
  {}
  [\rd_m - B_n,\, \rd_n - B_n] = 0, \quad 
  [\bar{\rd}_m - \bar{B}_n,\, \bar{\rd}_n - \bar{B}_n] = 0,\\
  [\rd_m - B_m,\, \bar{\rd}_n - \bar{B}_n] = 0. 
\end{gathered}
\label{Toda-ZS}
\eeq
We can construct two Lax operators of the form 
\beqnn
  L = e^{\rd_s} + \sum_{j=0}^\infty u_je^{-j\rd_s}, \quad 
  \bar{L} = \sum_{j=1}^\infty \bar{u}_je^{j\rd_s}  
\eeqnn
such that the Zakharov-Shabat operators 
are expressed as 
\beqnn
  B_n = (L^n)_{\ge 0}, \quad 
  \bar{B}_n = (\bar{L}^n)_{<0}, 
\eeqnn
where $(\quad)_{\ge 0}$ and $(\quad)_{<0}$ 
are the projection onto nonnegative and 
negative powers of $e^{\rd_s}$, respectively.  
The Lax operators satisfy the Lax equations 
\beq
\begin{aligned}
  \rd_n(L) &= [B_n,L], \quad 
  \rd_n(\bar{L}) &= [B_n,\bar{L}], \\
  \bar{\rd}_n(L) &= [\bar{B}_n,L], \quad 
  \bar{\rd}_n(\bar{L}) &= [\bar{B}_n,\bar{L}]. 
\end{aligned}
\label{Toda-Lax}
\eeq

\subsection{Difference Fay identities and auxiliary linear equations}

We now rewrite the difference Fay identities 
(\ref{Toda-diffFay1(i)}) -- (\ref{Toda-diffFay1(iii)}) 
in the language of the wave functions.  
This leads to a generating functional 
expression of the auxiliary linear problem. 

Let us first consider (\ref{Toda-diffFay1(i)}). 
We shift the time variables as 
$\bst \to \bst-[\lambda^{-1}]-[\mu^{-1}]$ 
and multiply both hand sides by 
$(\lambda-\mu)e^{\xi(\bst,\mu)}/
\tau(s+1,\bst,\bar{\bst})\tau(s,\bst-[\lambda^{-1}],\bar{\bst})$. 
This yields the equation 
\begin{multline*}
  (\lambda-\mu)e^{\xi(\bst,\mu)}e^{-D(\lambda)} 
   \frac{\tau(s,\bst-[\mu^{-1}],\bar{\bst})}
        {\tau(s,\bst,\bar{\bst})} \\
   = \lambda
     \frac{\tau(s+1,\bst-[\lambda^{-1}],\bar{\bst})/ 
           \tau(s+1,\bst,\bar{\bst})} 
          {\tau(s,\bst-[\lambda^{-1}],\bar{\bst})/ 
           \tau(s,\bst,\bar{\bst})} 
     \mu^se^{\xi(\bst,\mu)} 
     \frac{\tau(s,\bst-[\mu^{-1}],\bar{\bst})}
          {\tau(s,\bst,\bar{\bst})} \\
   - \mu^{s+1}e^{\xi(\bst,\mu)} 
     \frac{\tau(s+1,\bst-[\mu^{-1}],\bar{\bst})} 
          {\tau(s,\bst,\bar{\bst})}. 
\end{multline*}
Noting that the commutator of $e^{-D(\lambda)}$ 
and $e^{\xi(\bst,\mu)}$ yields the factor 
$1 - \mu/\lambda$, we can rewrite 
the last equation as 
\beq
  \lambda e^{-D(\lambda)}\Psi(s,\bst,\bar{\bst},\mu)
  = \frac{\Psi(s+1,\bst,\bar{\bst},\lambda)}
         {\Psi(s,\bst,\bar{\bst},\lambda)}
    \Psi(s,\bst,\bar{\bst},\mu) 
  - \Psi(s+1,\bst,\bar{\bst},\mu). 
\label{Toda-auxlin-gen1}
\eeq

Let us now consider (\ref{Toda-diffFay1(ii)}). 
We now shift $\bar{\bst}$ and $s$ as 
$\bar{\bst} \to \bar{\bst}-[\lambda]-[\mu]$ 
and $s \to s+2$, and multiply both hand side 
by $\lambda(\lambda-\mu)\mu^se^{\xi(\bar{\bst},\mu^{-1})}/ 
\tau(s,\bst,\bar{\bst})\tau(s,\bst,\bar{\bst}-[\lambda])$. 
This yields the equation 
\begin{multline*}
   - \frac{\lambda-\mu}{\mu}\mu^se^{\xi(\bar{\bst},\mu^{-1})} 
     e^{-\bar{D}(\lambda)}
     \frac{\tau(s+1,\bst,\bar{\bst}-[\mu])}
          {\tau(s,\bst,\bar{\bst})} \\
  = \mu^se^{\xi(\bar{\bst},\mu^{-1})} 
    \frac{\tau(s+1,\bst,\bar{\bst}-[\mu^{-1}])} 
         {\tau(s,\bst,\bar{\bst})} 
  - \lambda 
    \frac{\tau(s+1,\bst,\bar{\bst}-[\lambda])/ 
          \tau(s,\bst,\bar{\bst})} 
         {\tau(s,\bst,\bar{\bst}-[\lambda])/ 
          \tau(s-1,\bst,\bar{\bst})} 
    \times \\
    \times \mu^{s-1}e^{\xi(\bst,\mu)} 
    \frac{\tau(s,\bst,\bar{\bst}-[\mu])} 
          {\tau(s-1,\bst,\bar{\bst})}, 
\end{multline*}
which can be cast into the form 
\beq
  e^{-\bar{D}(\lambda)}\bar{\Psi}(s,\bst,\bar{\bst},\mu)
  = \bar{\Psi}(s,\bst,\bar{\bst},\mu) 
  - \frac{\bar{\Psi}(s,\bst,\bar{\bst},\lambda)} 
         {\bar{\Psi}(s-1,\bst,\bar{\bst},\lambda)} 
    \bar{\Psi}(s-1,\bst,\bar{\bst},\mu). 
\label{Toda-auxlin-gen2}
\eeq

In the same way, we can derive 
two more linear equations 
for the wave functions from 
(\ref{Toda-diffFay(iii)}).  
Firstly, if we shift the time variables 
as $\bst \to \bst - [\lambda^{-1}]$ 
and $\bar{\bst} \to \bar{\bst} - [\mu]$ 
and multiply both hand side by 
$\lambda^se^{\xi(\bst,\lambda)}/ 
\tau(s,\bst,\bar{\bst})\tau(s,\bst,\bar{\bst}-[\mu])$, 
we obtain the equation 
\beqnn
  e^{-\bar{D}(\mu)}\Psi(s,\bst,\bar{\bst},\lambda) 
  = \Psi(s,\bst,\bar{\bst},\lambda) 
  - \frac{\bar{\Psi}(s,\bst,\bar{\bst},\mu)} 
         {\bar{\Psi}(s-1,\bst,\bar{\bst},\mu)} 
    \Psi(s-1,\bst,\bar{\bst},\lambda). 
\eeqnn
It will be better to exchange $\lambda$ and $\mu$ as 
\beq
  e^{-\bar{D}(\lambda)}\Psi(s,\bst,\bar{\bst},\mu)
  = \Psi(s,\bst,\bar{\bst},\mu) 
  - \frac{\bar{\Psi}(s,\bst,\bar{\bst},\lambda)} 
         {\bar{\Psi}(s-1,\bst,\bar{\bst},\lambda)} 
    \Psi(s-1,\bst,\bar{\bst},\mu), 
\label{Toda-auxlin-gen3}
\eeq
because the outcome takes the same form 
as (\ref{Toda-auxlin-gen2}). 
Secondly, if we shift $\bst,\bar{\bst}$ and 
$s$ as $\bst \to \bst-[\lambda^{-1}]$, 
$\bar{\bst} \to \bar{\bst}-[\mu]$, 
$s \to s+1$ and multiply both hand side by 
$\lambda\mu^se^{\xi(\bar{\bst},\mu^{-1})}/
\tau(s+1,\bst,\bar{\bst})\tau(s,\bst-[\lambda^{-1}],\bar{\bst})$, 
we are led to the equation 
\beq
  \lambda e^{-D(\lambda)}\bar{\Psi}(s,\bst,\bar{\bst},\mu)
  = \frac{\Psi(s+1,\bst,\bar{\bst},\lambda)} 
         {\Psi(s,\bst,\bar{\bst},\lambda)} 
    \bar{\Psi}(s,\bst,\bar{\bst},\mu) 
  - \bar{\Psi}(s+1,\bst,\bar{\bst},\mu). 
\label{Toda-auxlin-gen4}
\eeq

(\ref{Toda-auxlin-gen1}) -- (\ref{Toda-auxlin-gen4}) 
show that the linear equations 
\beqnn
\begin{gathered}
  \lambda e^{-D(\lambda)}\Phi 
  = \frac{\Psi(s+1,\bst,\bar{\bst},\lambda)} 
    {\Psi(s,\bst,\bar{\bst},\lambda)}\Phi 
  - e^{\rd_s}\Phi, \\  
  e^{-\bar{D}(\lambda)}\Phi 
  = \Phi 
  - \frac{\bar{\Psi}(s,\bst,\bar{\bst},\lambda)} 
    {\bar{\Psi}(s-1,\bst,\bar{\bst},\lambda)}
    e^{-\rd_s}\Phi 
\end{gathered}
\eeqnn
are satisfied by both 
$\Phi = \Psi(s,\bst,\bar{\bst},\mu)$ 
and $\Phi = \bar{\Psi}(s,\bst,\bar{\bst},\mu)$. 
Expanded in powers of $\lambda$, 
they generate an infinite number 
of linear equations of the form 
\beq
  h_n(-\tilde{\rd}_{\bst})\Phi = v_n\Phi,  \quad 
  h_n(-\tilde{\rd}_{\bar{\bst}})\Phi 
  = - \bar{v}_ne^{-\rd_s}\Phi, \quad 
\label{Toda-auxlin-SS1}
\eeq
where $\tilde{\rd}_{\bst}$ is the same notation 
as defined for the KP hierarchy, 
$\tilde{\rd}_{\bar{\bst}}$ is similarly defined as 
\beqnn
  \tilde{\rd}_{\bar{\bst}} 
  = \Bigl(\bar{\rd}_1,\frac{\bar{\rd}_2}{2},\ldots,
          \frac{\bar{\rd}_n}{n},\ldots\Bigr), 
\eeqnn
and $v_j = v_j(s,\bst,\bar{\bst})$ and 
$\bar{v}_j = \bar{v}_j(s,\bst,\bar{\bst})$ 
are the Laurent coefficients of 
the $\Psi$-quotients in the foregoing 
linear equations, namely, 
\beqnn
  \frac{\Psi(s+1,\bst,\bar{\bst},\lambda)}
       {\Psi(s,\bst,\bar{\bst},\lambda)} 
  = \lambda + \sum_{n=1}^\infty v_n\lambda^{1-n}, \quad 
  \frac{\bar{\Psi}(s,\bst,\bar{\bst},\lambda)} 
       {\bar{\Psi}(s-1,\bst,\bar{\bst},\lambda)} 
  = \sum_{n=1}^\infty \bar{v}_n\lambda^n. 
\eeqnn
Note that the lowest ($n = 1$) equations of 
(\ref{Toda-auxlin-SS1}) read 
\beq
  \rd_1\Phi = (e^{\rd_s} - v_1)\Phi, \quad 
  \bar{\rd}_1\Phi = \bar{v}_1e^{-\rd_s}\Phi. 
\label{Toda-auxlin-SS2}
\eeq

As we have illustrated in the case of 
the KP hierarchy, one can derive 
the usual evolutionary form (\ref{Toda-auxlin}) 
of auxiliary linear equations from 
these equations (\ref{Toda-auxlin-SS1}) 
and (\ref{Toda-auxlin-SS2}). 
On the basis of this fact, Teo \cite{Teo06} proved 
that the difference Fay identities are equivalent 
to the full system of the Toda hierarchy.

\subsection{Dispersionless Hirota equations 
and Hamilton-Jacobi equations}

The quasi-classical ansatz (\ref{KP-tau-hbar}) 
for the tau function of the KP hierarchy 
can be readily generalized to 
the Toda hierarchy  \cite{TT93}.  
Namely, we allow the tau function 
to depend on $\hbar$ as $\tau 
= \tau(\hbar,s,\bst,\bar{\bst})$, 
and assume that the rescaled tau function 
$\tau_\hbar(s,\bst,\bar{\bst}) 
= \tau(\hbar,\hbar^{-1}s,\hbar^{-1}\bst,
\hbar^{-1}\bar{\bst})$ behave as 
\beq
  \tau_\hbar(s,\bst,\bar{\bst}) 
  = \exp\Bigl(\hbar^{-2}F(s,\bst,\bar{\bst}) 
              + O(\hbar^{-1})\Bigr)  
\label{Toda-tau-hbar}
\eeq
in the limit of $\hbar \to 0$. 
Note that the rescaling changes 
the lattice spacing from $1$ to $\hbar$, 
eventually tending to $0$ in 
the quasi-classical limit.  
The discrete variable $s$ in 
$\tau(\hbar,s,\bst,\bar{\bst})$ is thereby 
replaced by a continuous variable 
in $F(s,\bst,\bar{\bst})$.  

Under this quasi-classical ansatz, 
we can derive the following three 
dispersionless Hirota equations 
\cite{KKMWWZ01,Zabrodin01,BMRWZ01} 
for the $F$ function $F 
= F(s,\bst,\bar{\bst})$ from 
the difference Fay identities: 
\beq
\begin{gathered}
  e^{D(\lambda)D(\mu)F} 
  = \frac{\lambda e^{-D(\lambda)\rd_sF} 
       - \mu e^{-D(\mu)\rd_sF}}{\lambda - \mu},\\
  e^{\bar{D}(\lambda)\bar{D}(\mu)F} 
  = \frac{-\mu e^{\bar{D}(\lambda)\rd_sF} 
       + \lambda e^{\bar{D}(\mu)\rd_sF}}{\lambda - \mu},\\
  e^{D(\lambda)\bar{D}(\mu)F} 
  = 1 - \frac{\mu}{\lambda} 
         e^{(D(\lambda)-\bar{D}(\mu)+\rd_s)\rd_sF}. 
\end{gathered}
\label{dToda-Hirota1}
\eeq
It will be better to rewrite the second equation as 
\beqnn
  e^{\bar{D}(\lambda)\bar{D}(\mu)F} 
  = \frac{\lambda^{-1}e^{\bar{D}(\lambda)\rd_sF} 
      - \mu^{-1}e^{\bar{D}(\mu)\rd_sF}}
    {\lambda^{-1} - \mu^{-1}}, 
\eeqnn
by which the symmetry between $\bst$ 
and $\bar{\bst}$ becomes manifest.  
Moreover, defining the $S$ functions as 
\beqnn
\begin{gathered}
  S(z) = \xi(\bst,z) + s\log z - D(z)F, \\
  \bar{S}(z) = \xi(\bar{\bst},z^{-1}) + s\log z
     + \rd_sF - \bar{D}(z)F, 
\end{gathered}
\eeqnn
we can rewrite (\ref{dToda-Hirota1}) as 
\beq
\begin{gathered}
  e^{D(\lambda)D(\mu)F} 
  = \frac{e^{\rd_sS(\lambda)} 
         - e^{\rd_sS(\mu)}}{\lambda-\mu}, \\
  e^{\bar{D}(\lambda)\bar{D}(\mu)F} 
  = e^{\rd_s^2F} 
     \frac{e^{-\rd_s\bar{S}(\lambda)} 
         - e^{-\rd_s\bar{S}(\mu)}}
     {\lambda^{-1} - \mu^{-1}}, \\
  e^{D(\lambda)\bar{D}(\mu)F} 
  = 1 - e^{-\rd_sS(\lambda)}
         e^{\rd_s\bar{S}(\mu)}. 
\end{gathered}
\label{dToda-Hirota2}
\eeq

The $S$ functions are actually the phase functions 
in the WKB ansatz 
\beq
\begin{gathered}
  \Psi_\hbar(s,\bst,\bar{\bst},z) 
  = \exp\Bigl(\hbar^{-1}S(z) + O(\hbar^0)\Bigr), \\
  \bar{\Psi}_\hbar(s,\bst,\bar{\bst},z) 
  = \exp\Bigl(\hbar^{-1}\bar{S}(z) + O(\hbar^0)\Bigr) 
\end{gathered}
\label{Toda-hbar-Psi}
\eeq
for the rescaled wave functions.  
The auxiliary linear equations thereby 
yield the Hamilton-Jacobi equations 
\beq
  \rd_nS = \calB_n(e^{\rd_sS}), \quad 
  \bar{\rd}_nS = \bar{\calB}_n(e^{\rd_sS})
\label{dToda-HJ}
\eeq
for $S = S(z), \bar{S}(z)$.  Here 
$\calB_n(P)$ and $\bar{\calB}_n(P)$ are 
classical counterparts of $B_n = B_n(e^{\rd_s})$ 
and $\bar{B}_n = \bar{B}_n(e^{\rd_s})$: 
\beqnn
  \calB_n(P) 
  = \lim_{\hbar\to 0}B_{\hbar,n}(P), \quad 
  \bar{\calB}_n(P) 
  = \lim_{\hbar\to 0}\bar{B}_{\hbar,n}(P).  
\eeqnn
The dispersionless Hirota equations 
in the form of (\ref{dToda-Hirota2}) 
turn out to be equivalent to these 
Hamilton-Jacobi equations, hence to 
the dispersionless Toda hierarchy itself 
\cite{BMRWZ01,BKMA03,Teo03}.

\section{BKP hierarchy and differential Fay identities}

\subsection{Bilinear equations for tau functions}

In this section, we mostly consider 
the two-component version 
\cite{DJKM-2BKP,KvdL98} of 
the BKP hierarchy \cite{DJKM-BKP}. 
This system contains two copies of 
the one-component BKP hierarchy as subsystems.  
The situation is thus somewhat similar to 
the Toda hierarchy, which may be thought of 
as the charged two-component KP hierarchy 
as well \cite{UT84}.  This is a reason 
why we are interested in the two-component case 
rather than the one-component BKP hierarchy.  

The two-component BKP hierarchy has 
two sets of time variables 
$\bst = (t_1,t_3,\ldots)$ and 
$\bar{\bst} = (\bar{t}_1,\bar{t}_3,\ldots)$ 
indexed by odd positive integers.  
The tau function $\tau = \tau(\bst,\bar{\bst})$ 
satisfies the bilinear equation 
\begin{multline}
   \oint\frac{dz}{2\pi iz}e^{\xi(\bst'-\bst,z)} 
   \tau(\bst'-2[z^{-1}],\bar{\bst}') 
   \tau(\bst+2[z^{-1}],\bar{\bst}) \\
 = \oint\frac{dz}{2\pi iz}e^{\xi(\bar{\bst}'-\bar{\bst},z)}
    \tau(\bst',\bar{\bst}'-2[z^{-1}]) 
    \tau(\bst,\bar{\bst}+2[z^{-1}]). 
\label{BKP-bilin-tau}
\end{multline}
The integrals on both hand sides 
are understood to be a contour integral 
along a sufficiently large circle $|z| = R$ 
(or just a formal algebraic operator).  
The other notations are as follows: 
\beqnn
  \xi(\bst,z) = \sum_{n=0}^\infty t_{2n+1}z^{2n+1}, \quad 
  [\alpha] = \Bigl(\alpha,\frac{\alpha^3}{3},\ldots,
             \frac{\alpha^{2n+1}}{2n+1},\ldots\Bigr). 
\eeqnn

If $\bst$ is specialized as $\bar{\bst}' = \bar{\bst}$, 
the bilinear equation (\ref{BKP-bilin-tau}) reduces to 
\beq
  \oint\frac{dz}{2\pi iz}e^{\xi(\bst'-\bst,z)}
  \tau(\bst'-2[z^{-1}],\bar{\bst}) 
  \tau(\bst+2[z^{-1}],\bar{\bst}) 
  = \tau(\bst',\bar{\bst})\tau(\bst,\bar{\bst}).  
\label{1BKP-bilin-tau1}
\eeq
Similarly, specializing $\bst'$ as 
$\bst' = \bst$ yields 
\beq
  \oint\frac{dz}{2\pi iz}e^{\xi(\bar{\bst}'-\bar{\bst},z)}
  \tau(\bst,\bar{\bst}'-2[z^{-1}]) 
  \tau(\bst,\bar{\bst}+2[z^{-1}]) 
  = \tau(\bst,\bar{\bst}')\tau(\bst,\bar{\bst}). 
\label{1BKP-bilin-tau2}
\eeq
These equations coincide with the bilinear equation 
of the one-component BKP hierarchy.  Thus 
the tau function of the two-component 
BKP hierarchy may be thought of as 
the tau function of the one-component 
BKP hierarchy with respect to 
both $\bst$ and $\bar{\bst}$.  

(\ref{BKP-bilin-tau}) can be converted 
to the Hirota form 
\begin{multline}
    \sum_{n=0}^\infty 
    h_n(-2\bsa)h_n(2\tilde{D}_{\bst})
    e^{\langle\bsa,D_{\bst}\rangle 
       + \langle\bar{\bsa},D_{\bar{\bst}}\rangle} 
    \tau(\bst,\bar{\bst})\cdot\tau(\bst,\bar{\bst}) \\
  = \sum_{n=0}^\infty 
    h_n(-2\bar{\bsa})
    h_n(2\tilde{D}_{\bar{\bst}})
    e^{\langle\bsa,D_{\bst}\rangle 
       + \langle\bar{\bsa},D_{\bar{\bst}}\rangle} 
    \tau(\bst,\bar{\bst})\cdot\tau(\bst,\bar{\bst}), 
\label{BKP-Hirota-gen}
\end{multline}
where $\bsa = (a_1,a_3,\ldots)$ and 
$\bar{\bsa} = (\bar{a}_1,\bar{a}_3,\ldots)$ 
are new variables, $D_{\bst},\tilde{D}_{\bar{\bst}}$, 
and $\langle\bsa,D_{\bst}\rangle,
\langle\bar{\bsa},D_{\bar{\bst}}\rangle$ 
are operators defined as 
\beqnn
\begin{gathered}
  \tilde{D}_{\bst} 
  = \Bigl(D_1,\frac{D_3}{3},\ldots, 
          \frac{D_{2n+1}}{2n+1},\ldots\Bigr), \quad
  \langle\bsa,D_{\bst}\rangle 
  = \sum_{n=0}^\infty a_{2n+1}D_{2n+1}, \\
  \tilde{D}_{\bar{\bst}} 
  = \Bigl(\bar{D}_1,\frac{\bar{D}_3}{3},\ldots, 
          \frac{\bar{D}_{2n+1}}{2n+1},\ldots\Bigr), \quad
  \langle\bar{\bsa},\bar{D}_{\bar{\bst}}\rangle 
  = \sum_{n=0}^\infty \bar{a}_{2n+1}\bar{D}_{2n+1},  
\end{gathered}
\eeqnn
and $h_n(\bst)$ are defined by the generating 
function 
\beqnn
  e^{\xi(\bst,z)} 
  = \exp\Bigl(\sum_{n=0}^\infty t_{2n+1}z^{2n+1}\Bigr)  
  = \sum_{n=0}^\infty h_n(\bst)z^n. 
\eeqnn

\subsection{Differential Fay identities}

We can derive four differential Fay identities 
\cite{Takasaki06} from (\ref{BKP-bilin-tau}) 
as follows. 
\begin{itemize}
\item[(i)] Differentiate the bilinear equation 
by $t'_1$ and specialize $\bst',\bar{\bst}'$ 
as $\bst' = \bst+2[\lambda^{-1}]+2[\mu^{-1}]$, 
$\bar{\bst}' = \bar{\bst}$.  
\item[(ii)] Differentiate the bilinear equation 
by $\bar{t}'_1$ and specialize $\bst',\bar{\bst}'$ 
as $\bst' = \bst$, $\bar{\bst}' 
= \bar{\bst}+2[\lambda^{-1}]+2[\mu^{-1}]$. 
\item[(iii)] Differentiate the bilinear equation 
by $t'_1$ and specialize $\bst,\bar{\bst}'$ 
as $\bst' = \bst+2[\lambda^{-1}]$, 
$\bar{\bst}' = \bar{\bst}+2[\mu^{-1}]$. 
\item[(iv)] Differentiate the bilinear equation 
by $\bar{t}'_1$ and specialize $\bst,\bar{\bst}'$ 
as $\bst' = \bst+2[\lambda^{-1}]$, 
$\bar{\bst}' = \bar{\bst}+2[\mu^{-1}]$. 
\end{itemize}
Note that the exponential factors 
in the bilinear equations become 
rational functions by virtue of the identity 
\beqnn
  \sum_{n=0}^\infty 
  \frac{2}{2n+1}\Bigl(\frac{z}{w}\Bigr)^{2n+1}
  = - \log\frac{1 - z/w}{1 + z/w}. 
\eeqnn
For example, in the case of (i), 
\beqnn
  e^{\xi(\bst'-\bst,z)} 
  = \frac{(z+\lambda)(z+\mu)}{(z-\lambda)(z-\mu)}, \quad 
  e^{\xi(\bar{\bst}'-\bar{\bst},z)} 
  = 1. 
\eeqnn
We can thus calculate the contour integrals 
by residue calculus. 

\paragraph*{(i)} 
After differentiating by $t'_1$ and 
specializing $\bst'$, the bilinear equation becomes 
\begin{multline*}
 \oint\frac{dz}{2\pi iz}
 \frac{(z+\lambda)(z+\mu)}{(z-\lambda)(z-\mu)} 
 \Bigl(z\tau(\bst+2[\lambda^{-1}]+2[\mu^{-1}]-2[z^{-1}],\bar{\bst}) \\
   + (\rd_1\tau)(\bst+2[\lambda^{-1}]+2[\mu^{-1}],\bar{\bst})
 \Bigr)\tau(\bst+2[z^{-1}],\bar{\bst}) \\
 = \oint\frac{dz}{2\pi iz} 
   (\rd_1\tau)(\bst+2[\lambda^{-1}]+2[\mu^{-1}],\bar{\bst}-2[z^{-1}]) 
   \tau(\bst,\bar{\bst}+2[z^{-1}]). 
\end{multline*}
By residue calculus, we obtain 
the first differential Fay identity  
\begin{multline}
  \left(\lambda+\mu 
    - \rd_1\log
      \frac{\tau(\bst+2[\lambda^{-1}]+2[\mu^{-1}],\bar{\bst})}
           {\tau(\bst,\bar{\bst})}\right) 
  \times\\
  \times 
  \frac{\tau(\bst+2[\lambda^{-1}]+2[\mu^{-1}],\bar{\bst}) 
        \tau(\bst,\bar{\bst})} 
       {\tau(\bst+2[\lambda^{-1}],\bar{\bst}) 
        \tau(\bst+2[\mu^{-1}],\bar{\bst})} \\
  = \frac{\lambda+\mu}{\lambda-\mu} 
    \left(\lambda-\mu
      - \rd_1\log
        \frac{\tau(\bst+2[\lambda^{-1}],\bar{\bst})}
             {\tau(\bst+2[\mu^{-1}],\bar{\bst})}\right). 
\label{BKP-diffFay1(i)}
\end{multline}
  
\paragraph*{(ii)} 
This case is substantially the same as (i), 
only the role of $\bst$ and $\bar{\bst}$ 
being exchanged.  We thus obtain the second 
differential Fay identity 
\begin{multline}
  \left(\lambda+\mu 
    - \bar{\rd}_1
      \log\frac{\tau(\bst,\bar{\bst}+2[\lambda^{-1}]+2[\mu^{-1}])}
               {\tau(\bst,\bar{\bst})}\right) 
  \times\\
  \times
  \frac{\tau(\bst,\bar{\bst}+2[\lambda^{-1}]+2[\mu^{-1}]) 
        \tau(\bst,\bar{\bst})} 
       {\tau(\bst,\bar{\bst}+2[\lambda^{-1}]) 
        \tau(\bst,\bar{\bst}+2[\mu^{-1}])} \\
  = \frac{\lambda+\mu}{\lambda-\mu} 
    \left(\lambda-\mu
      - \bar{\rd}_1\log
        \frac{\tau(\bst,\bar{\bst}+2[\lambda^{-1}])}
             {\tau(\bst,\bar{\bst}+2[\mu^{-1}])}\right). 
\label{BKP-diffFay1(ii)}
\end{multline}

\paragraph*{(iii)} 
After differentiation and specialization, 
the bilinear equation becomes 
\begin{multline*}
 \oint\frac{dz}{2\pi iz}
 \frac{(z+\lambda)(z+\mu)}{(z-\lambda)(z-\mu)}
 \Bigl(z\tau(\bst+2[\lambda^{-1}]-2[z^{-1}],
             \bar{\bst}+2[\mu^{-1}]) \\
  + (\rd_1\tau)(\bst+2[\lambda^{-1}]-2[z^{-1}],
                       \bar{\bst}+2[\mu^{-1}])
 \Bigr)\tau(\bst+2[z^{-1}],\bar{\bst}) \\
 = \oint\frac{dz}{2\pi iz}
   \frac{z+\mu}{z-\mu}
   (\rd_1\tau)(\bst+2[\lambda^{-1}],\bar{\bst}+2[\mu^{-1}]-2[z^{-1}])
   \tau(\bst,\bar{\bst}+2[z^{-1}]). 
\end{multline*}
By residue calculus, we obtain the third 
differential Fay identity
\begin{multline}
  \left(\lambda 
    - \rd_1\log\
      \frac{\tau(\bst+2[\lambda^{-1}],\bar{\bst}+2[\mu^{-1}])}
           {\tau(\bst,\bar{\bst})}\right) 
  \times\\
  \times
  \frac{\tau(\bst+2[\lambda^{-1}],\bar{\bst}+2[\mu^{-1}])
        \tau(\bst,\bar{\bst})} 
       {\tau(\bst+2[\lambda^{-1}],\bar{\bst})
        \tau(\bst,\bar{\bst}+2[\mu^{-1}])} \\
  = \lambda 
    - \rd_1\log\frac{\tau(\bst+2[\lambda^{-1}],\bar{\bst})}
                    {\tau(\bst,\bar{\bst}+[\mu^{-1}])}. 
\label{BKP-diffFay1(iii)}
\end{multline}

\paragraph*{(iv)} 
This case is parallel to (iii), and leads to 
the fourth differential Fay identity 
\begin{multline}
  \left(\mu 
    - \bar{\rd}_1\log
      \frac{\tau(\bst+2[\lambda^{-1}],\bar{\bst}+[\mu^{-1}])}
           {\tau(\bst,\bar{\bst})} \right) 
  \times\\
  \times
  \frac{\tau(\bst+2[\lambda^{-1}],\bar{\bst}+2[\mu^{-1}])
        \tau(\bst,\bar{\bst})}
       {\tau(\bst+2[\lambda^{-1}],\bar{\bst})
        \tau(\bst,\bar{\bst}+2[\mu^{-1}])} \\
  = \mu 
    - \bar{\rd}_1\log\frac{\tau(\bst,\bar{\bst}+2[\mu^{-1}])}
                          {\tau(\bst+2[\lambda^{-1}],\bar{\bst})}. 
\label{BKP-diffFay1(iv)}
\end{multline}

We can use the BKP version 
\beqnn
  D(z) = \sum_{n=0}^\infty \frac{z^{-2n-1}}{2n+1}\rd_{2n+1}, 
  \quad 
  \bar{D}(z) = \sum_{n=0}^\infty \frac{z^{-2n-1}}{2n+1}\bar{\rd}_{2n+1}
\eeqnn
of the $D(z)$ operator to rewrite 
the foregoing differential Fay identities 
as follows: 
\begin{multline}
  \Bigl(\lambda + \mu 
    - \rd_1(e^{2D(\lambda)+2D(\mu)}-1)\log\tau\Bigr) 
  \times\\
  \times
  \exp\Bigl((e^{2D(\lambda)}-1)(e^{2D(\mu)}-1)\log\tau\Bigr)\\
  = \frac{\lambda+\mu}{\lambda-\mu} 
    \Bigl(\lambda - \mu 
      - \rd_1(e^{2D(\lambda)}-e^{2D(\mu)})\log\tau\Bigr), 
\label{BKP-diffFay2(i)}
\end{multline}
\begin{multline}
  \Bigl(\lambda + \mu 
    - \bar{\rd}_1(e^{2\bar{D}(\lambda)+2\bar{D}(\mu)}-1)\log\tau\Bigr)
  \times\\
  \times
  \exp\Bigl((e^{2\bar{D}(\lambda)}-1)(e^{2\bar{D}(\mu)}-1)\log\tau\Bigr)\\
  = \frac{\lambda+\mu}{\lambda-\mu} 
    \Bigl(\lambda - \mu 
      - \bar{\rd}_1(e^{2\bar{D}(\lambda)}-e^{2\bar{D}(\mu)})\log\tau\Bigr), 
\label{BKP-diffFay2(ii)}
\end{multline}
\begin{multline}
  \Bigl(\lambda 
    - \rd_1(e^{2D(\lambda)+2\bar{D}(\mu)}-1)\log\tau\Bigr) 
  \times\\
  \times
  \exp\Bigl((e^{2D(\lambda)}-1)(e^{2\bar{D}(\mu)}-1)\log\tau\Bigr)\\
  = \lambda - \rd_1(e^{2D(\lambda)}-e^{2\bar{D}(\mu)})\log\tau, 
\label{BKP-diffFay2(iii)}
\end{multline}  
\begin{multline}
  \Bigl(\mu 
    - \bar{\rd}_1(e^{2D(\lambda)+2\bar{D}(\mu)}-1)\log\tau\Bigr) 
  \times\\
  \times
  \exp\Bigl((e^{2D(\lambda)}-1)(e^{2\bar{D}(\mu)}-1)\log\tau\Bigr)\\
  = \mu - \bar{\rd}_1(e^{2\bar{D}(\mu)}-e^{2D(\lambda)})\log\tau. 
\label{BKP-diffFay2(iv)}
\end{multline}

\subsection{Auxiliary linear equations}

We now introduce the two wave functions
\beqnn
\begin{gathered}
  \Psi(\bst,\bar{\bst},z) 
  = \frac{\tau(\bst-2[z^{-1}],\bar{\bst})}
    {\tau(\bst,\bar{\bst})}e^{\xi(\bst,z)}, \\
  \bar{\Psi}(\bst,\bar{\bst},z) 
  = \frac{\tau(\bst,\bar{\bst}-2[z^{-1}])}
    {\tau(\bst,\bar{\bst})}e^{\xi(\bar{\bst},z)}. 
\end{gathered}
\eeqnn
Their duals are given by $\Psi(\bst,\bar{\bst},-z)$ 
and $\bar{\Psi}(\bst,\bar{\bst},-z)$.  
The bilinear equation (\ref{BKP-bilin-tau}) 
for the tau function yields the bilinear equation 
\beq
  \oint\frac{dz}{2\pi iz}
  \Psi(\bst',\bar{\bst}',z)\Psi(\bst,\bar{\bst},-z) 
= \oint\frac{dz}{2\pi iz}
  \bar{\Psi}(\bst',\bar{\bst}',z)\bar{\Psi}(\bst,\bar{\bst},-z) 
\label{BKP-bilin-Psi}
\eeq
for these wave functions.  From this bilinear 
equations, one can derive an infinite 
number of auxiliary linear equations of the form 
\beq
  (\rd_{2n+1} - B_{2n+1})\Phi = 0, \quad 
  (\bar{\rd}_{2n+1} - \bar{B}_{2n+1})\Phi = 0 
\label{BKP-auxlin1}
\eeq
for $n = 0,1,\ldots$ and 
\beq
  (\rd_1\bar{\rd}_1 - u)\Phi = 0 
\label{BKP-auxlin2}
\eeq
that hold for both $\Phi = \Psi(\bst,\bar{\bst},z)$ 
and $\Phi = \bar{\Psi}(\bst,\bar{\bst},z)$. 
$B_{2n+1} = B_{2n+1}(\rd_1)$ and 
$\bar{B}_{2n+1} = \bar{B}_{2n+1}(\bar{\rd}_1)$ 
are differential operators with respect 
to $t_1$ and $\bar{t}_1$ of the form 
\beqnn
\begin{gathered}
  B_{2n+1} 
  = \rd_1^{2n+1} + b_{2n+1,2}\rd_1^{2n-1} 
     + \cdots + b_{2n+1,2n}\rd_1, \\
  \bar{B}_{1n+1}
  = \bar{\rd}_1^{2n+1} + \bar{b}_{2n+1,2}\bar{\rd}_1^{2n-1} 
     + \cdots + \bar{b}_{2n+1,2n}\bar{\rd}_1. 
\end{gathered}
\eeqnn
Note that they do not have a $0$-th order term.  
This is a characteristic of the operators 
that show up in the auxiliary linear problem 
of the one-component BKP hierarchy \cite{DJKM-BKP}. 
The first two sets of equations (\ref{BKP-auxlin1}) 
are thus nothing but the auxiliary linear problem 
of the underlying one-component BKP hierarchies.  
The third equation (\ref{BKP-auxlin2}) may be 
thought of as the two-dimensional ``integrable 
Schr\"odinger equation'' associated with 
the Novikov-Veselov equation \cite{NV84}.  
The potential $u$ is given by 
\beq
  u = - 2\rd_1\bar{\rd}_1\log\tau. 
\label{BKP-u}
\eeq
In this respect, the two-component BKP hierarchy 
is also referred to as the Novikov-Veselov 
hierarchy.  $t_1$ and $\bar{t}_1$ play 
the role of spatial variables therein.  
Auxiliary linear equations of this type 
play a central role in Krichever's recent work 
\cite{Krichever05} on a Schottky-type problem 
on Prym varieties \cite{Shiota89}.  

The Lax formalism of this hierarchy 
exhibits new features because of the presence 
of two spatial dimensions.  For example, 
the Zakharov-Shabat operators of the same type 
satisfy the usual zero-curvature equations 
\beq
\begin{gathered}
  {}
  [\rd_{2m+1} - B_{2m+1},\, \rd_{2n+1} - B_{2n+1}] = 0, \\
  [\bar{\rd}_{2m+1} - \bar{B}_{2m+1},\, 
   \bar{\rd}_{2n+1} - \bar{B}_{2n+1}]  = 0, \\
\end{gathered}
\label{BKP-ZS1}
\eeq
but the equation for the pair of $B_{2m+1}$ 
and $\bar{B}_{2n+1}$ has an extra term 
on the right hand side as 
\beq
  [\rd_{2m+1} - B_{2m+1},\, 
   \bar{\rd}_{2n+1} - \bar{B}_{2n+1}] 
  = D_{mn}(\rd_1,\bar{\rd}_1)(\rd_1\bar{\rd}_1 - u), 
\label{BKP-ZS2}
\eeq
where $D_{mn}(\rd_1,\bar{\rd}_1)$ 
is a differential operator with respect 
to both $t_1$ and $\bar{t}_1$ 
\cite{NV84,Krichever05}.

\subsection{Differential Fay identities and auxiliary linear equations}

As we have demonstrated for the cases 
of the KP and Toda hierarchies, 
the differential Fay identities 
in the present case, too, give 
a generating functional expression 
of the auxiliary linear problem.  
The situation, however, turns out to be 
more complicated than the previous two cases.  

Let us first consider (\ref{BKP-diffFay1(i)}) 
and (\ref{BKP-diffFay1(ii)}).  
As regards (\ref{BKP-diffFay1(i)}), 
we first shift $\bst$ as $\bst \to 
\bst-2[\lambda^{-1}]-2[\mu^{-1}]$.  
The equation thereby changes to 
\begin{multline*}
  \left(\lambda+\mu 
    + \rd_1\log
      \frac{\tau(\bst-2[\lambda^{-1}]-2[\mu^{-1}],\bar{\bst})}
           {\tau(\bst,\bar{\bst})}\right) 
  \times\\
  \times
  \frac{\tau(\bst-2[\lambda^{-1}]-2[\mu^{-1}],\bar{\bst}) 
        \tau(\bst,\bar{\bst})} 
       {\tau(\bst-2[\lambda^{-1}],\bar{\bst}) 
        \tau(\bst-2[\mu^{-1}],\bar{\bst})} \\
  = \frac{\lambda+\mu}{\lambda-\mu} 
    \left(\lambda-\mu
      + \rd_1\log
        \frac{\tau(\bst-2[\lambda^{-1}],\bar{\bst})}
             {\tau(\bst-2[\mu^{-1}],\bar{\bst})}\right). 
\end{multline*}
We then multiply both hand sides by 
$e^{\xi(\bst,\mu)}\tau(\bst-2[\mu^{-1}],\bar{\bst})/
\tau(\bst,\bar{\bst})$.  After some algebra, 
we obtain the equation 
\begin{multline}
  (\rd_1\log\Psi(\bst,\bar{\bst},\lambda) + \rd_1)
  e^{-2D(\lambda)}\Psi(\bst,\bar{\bst},\mu) \\
= (\rd_1\log\Psi(\bst,\bar{\bst},\lambda) - \rd_1)
  \Psi(\bst,\bar{\bst},\mu). 
\label{BKP-auxlin-gen1}
\end{multline}
In much the same way, shifting $\bar{\bst}$ as 
$\bar{\bst} \to \bar{\bst}-2[\lambda^{-1}]-2[\mu^{-1}]$ 
in (\ref{BKP-diffFay1(ii)}) and multiplying it by 
$e^{\xi(\bar{\bst},\mu)}\tau(\bst,\bar{\bst}-2[\mu^{-1}])/
\tau(\bst,\bar{\bst})$, we obtain the equation 
\begin{multline}
  (\bar{\rd}_1\log\bar{\Psi}(\bst,\bar{\bst},\lambda) 
   + \bar{\rd}_1)e^{-2\bar{D}(\lambda)}
  \bar{\Psi}(\bst,\bar{\bst},\mu) \\
= (\bar{\rd}_1\log\bar{\Psi}(\bst,\bar{\bst},\lambda) 
   - \bar{\rd}_1)\bar{\Psi}(\bst,\bar{\bst},\mu). 
\label{BKP-auxlin-gen2}
\end{multline}

We can rewrite (\ref{BKP-diffFay1(iii)}) 
and (\ref{BKP-diffFay1(iv)}) in a similar way.  
As regards (\ref{BKP-diffFay1(iii)}), 
we shift both $\bst$ and $\bar{\bst}$ 
as $\bst \to \bst-2[\lambda^{-1}]$ 
and $\bar{\bst} \to \bar{\bst}-2[\mu^{-1}]$, 
and multiply both hand sides by 
$e^{\xi(\bar{\bst},\mu)}\tau(\bst,\bar{\bst}-2[\mu^{-1}])/
\tau(\bst,\bar{\bst})$.  This leads to  
the equation 
\begin{multline}
  (\rd_1\log\Psi(\bst,\bar{\bst},\lambda) + \rd_1)
  e^{-2D(\lambda)}\bar{\Psi}(\bst,\bar{\bst},\mu) \\
= (\rd_1\log\Psi(\bst,\bar{\bst},\lambda) - \rd_1)
  \bar{\Psi}(\bst,\bar{\bst},\mu). 
\label{BKP-auxlin-gen3}
\end{multline}
Note that this equation has the same form 
as (\ref{BKP-diffFay1(i)}).  
Anticipating a similar result, 
we now consider (\ref{BKP-diffFay1(iv)}). 
If we shift $\bst$ and $\bar{\bst}$ as 
$\bst \to \bst-2[\lambda^{-1}]$ and 
$\bar{\bst} \to \bar{\bst}-2[\mu^{-1}]$ 
and multiply this equation by 
$e^{\xi(\bst,\lambda)}\tau(\bst-2[\lambda^{-1}],\bar{\bst})/
\tau(\bst,\bar{\bst})$, the outcome 
is the equation 
\beqnn
  (\bar{\rd}_1\log\bar{\Psi}(\bst,\bar{\bst},\mu) 
   + \bar{\rd}_1)
  e^{-2\bar{D}(\mu)}\Psi(\bst,\bar{\bst},\lambda) 
= (\bar{\rd}_1\log\bar{\Psi}(\bst,\bar{\bst},\mu)
   - \bar{\rd}_1)\Psi(\bst,\bar{\bst},\lambda). 
\eeqnn
which is almost what we want. 
By exchanging $\lambda$ and $\mu$, 
we obtain the equation 
\begin{multline}
  (\bar{\rd}_1\log\bar{\Psi}(\bst,\bar{\bst},\lambda) 
   + \bar{\rd}_1)
  e^{-2\bar{D}(\lambda)}\Psi(\bst,\bar{\bst},\mu) \\
= (\bar{\rd}_1\log\bar{\Psi}(\bst,\bar{\bst},\lambda) 
   - \bar{\rd}_1)\Psi(\bst,\bar{\bst},\mu).  
\label{BKP-auxlin-gen4}
\end{multline}
As anticipated, this equation has the same form 
as (\ref{BKP-auxlin-gen2}).  

These results show that 
$\Psi(\bst,\bar{\bst},\mu)$ 
and $\bar{\Psi}(\bst,\bar{\bst},\mu)$ 
satisfy the same linear equations 
\beqnn
\begin{gathered}
  (\rd_1\log\Psi(\bst,\bar{\bst},\lambda) 
   + \rd_1)e^{-2D(\lambda)}\Phi 
  = (\rd_1\log\Psi(\bst,\bar{\bst},\lambda) 
      - \rd_1)\Phi, \\
  (\bar{\rd}_1\log\bar{\Psi}(\bst,\bar{\bst},\mu) 
   + \bar{\rd}_1)e^{-2\bar{D}(\lambda)}\Phi 
  = (\bar{\rd}_1\log\bar{\Psi}(\bst,\bar{\bst},\mu)
      - \bar{\rd}_1)\Phi. 
\end{gathered}
\eeqnn
Expanded in powers of $\lambda$, they 
give rise to an infinite number of 
equations of the form 
\beq
\begin{gathered}
  \Bigl(h_{n+1}(-2\tilde{\rd}_{\bst}) 
    + \sum_{m=1}^{n-1}
      v_{m+1}h_{n-m}(-2\tilde{\rd}_{\bst}) 
    + \rd_1h_n(-2\tilde{\rd}_{\bst})
  \Bigr)\Phi = 0, \\
  \Bigl(h_{n+1}(-2\tilde{\rd}_{\bar{\bst}}) 
    + \sum_{m=1}^{n-1}
      \bar{v}_{m+1}h_{n-m}(-2\tilde{\rd}_{\bar{\bst}})
    + \bar{\rd}_1h_n(-2\tilde{\rd}_{\bar{\bst}})
  \Bigr)\Phi = 0
\end{gathered}
\label{BKP-auxlin-SS}
\eeq
for $n = 2,3,\ldots$, where 
$\tilde{\rd}_{\bst}$ and $\tilde{\rd}_{\bar{\bst}}$ 
denote the set of operators 
\beqnn
  \tilde{\rd}_{\bst} 
  = \Bigl(\rd_1,\frac{\rd_3}{3},\ldots,
          \frac{\rd_{2n+1}}{2n+1},\ldots\Bigr), \quad 
  \tilde{\rd}_{\bar{\bst}} 
  = \Bigl(\bar{\rd}_1,\frac{\bar{\rd}_3}{3},\ldots
          \frac{\bar{\rd}_{2n+1}}{2n+1},\ldots\Bigr), 
\eeqnn
and $v_n$'s and $\bar{v}_n$'s are defined by 
generating functions as 
\beqnn
\begin{gathered}
  \rd_1\log\Psi(\bst,\bar{\bst},\lambda) 
  = \lambda + \sum_{n=1}^\infty v_{n+1}\lambda^{-n},\\
  \bar{\rd}_1\bar{\Psi}(\bst,\bar{\bst},\lambda) 
  = \lambda + \sum_{n=1}^\infty\bar{v}_{n+1}\lambda^{-n}.
\end{gathered}
\eeqnn
For example, the equations for $n = 2$ 
in (\ref{BKP-auxlin-SS}) take the form 
\beqnn
\begin{gathered}
  \Bigl(-\frac{2}{3}\rd_3 + \frac{2}{3}\rd_1^3 
    + 4(\rd_1^2\log\tau)\rd_1\Bigr)\Phi = 0, \\
  \Bigl(-\frac{2}{3}\bar{\rd}_3 + \frac{2}{3}\bar{\rd}_1^3 
    + 4(\bar{\rd}_1^2\log\tau)\bar{\rd}_1\Bigr)\Phi = 0 
\end{gathered}
\eeqnn
or, equivalently, 
\beqnn
\begin{gathered}
  \bigl(\rd_3 - \rd_1^3 
    - 6(\rd_1^2\log\tau)\rd_1\big)\Phi = 0,\\
  \bigl(\bar{\rd}_3 - \bar{\rd}_1^3 
    - 6(\bar{\rd}_1^2\log\tau)\bar{\rd}_1\bigr)\Phi = 0, 
\end{gathered}
\eeqnn
which coincide with the lowest equations 
of (\ref{BKP-auxlin1}).  
The equations for $n = 2k$ in (\ref{BKP-auxlin-SS}) 
take such a form as 
\beqnn
\begin{gathered}
  \Bigl(-\frac{2}{2k+1}\rd_{2k+1} 
    - C_m(\rd_1,\ldots,\rd_{2k-1})\Bigl)\Phi = 0,\\
  \Bigl(-\frac{2}{2k+1}\bar{\rd}_{2k+1} 
    - \bar{C}_m(\bar{\rd}_1,\ldots,\bar{\rd}_{2k-1}\Bigr)\Phi = 0, 
\end{gathered}
\eeqnn
where $C_m(\rd_1,\ldots,\rd_{2k-1})$ and 
$\bar{C}_m(\bar{\rd}_1,\ldots,\bar{\rd}_{2k-1})$ 
are differential operators containing 
$\rd_1,\ldots,\rd_{2k-1}$ and 
$\bar{\rd}_1,\ldots,\bar{\rd}_{2k-1}$, 
respectively, but no $0$-th order term.  
As explained in the case of the KP hierarchy, 
one can convert these equations to 
the usual evolutionary form 
(\ref{BKP-auxlin1}).  

Still missing is the two-dimensional 
Schr\"odinger equation (\ref{BKP-auxlin2}).  
To derive this equation, 
we convert (\ref{BKP-diffFay1(iii)}) 
and (\ref{BKP-diffFay1(iv)}) 
to yet another form as follows.  
As regards (\ref{BKP-diffFay1(iii)}), 
we first shift $\bst$ and $\bar{\bst}$ 
as $\bst \to \bst-2[\lambda^{-1}]$ 
and $\bar{\bst} \to \bar{\bst}-2[\mu^{-1}]$, 
them multiply it by $e^{\xi(\bst,\lambda)}
\tau(\bst-2[\lambda^{-1}],\bar{\bst})/
\tau(\bst,\bar{\bst})$.  This yields 
the equation 
\beqnn
  (\rd_1 + \rd_1\log\bar{\Psi}(\bst,\bar{\bst},\mu))
  e^{-2\bar{D}(\mu)}\Psi(\bst,\bar{\bst},\lambda) 
  = (\rd_1 - \rd_1\log\bar{\Psi}(\bst,\bar{\bst},\mu))
    \Psi(\bst,\bar{\bst},\lambda). 
\eeqnn
Exchanging $\lambda$ and $\mu$, 
we eventually obtain the equation 
\begin{multline}
  (\rd_1 + \rd_1\log\bar{\Psi}(\bst,\bar{\bst},\lambda))
  e^{-2\bar{D}(\lambda)}\Psi(\bst,\bar{\bst},\mu) \\
= (\rd_1 - \rd_1\log\bar{\Psi}(\bst,\bar{\bst},\lambda))
  \Psi(\bst,\bar{\bst},\mu). 
\label{BKP-auxlin-gen5}
\end{multline}
Similarly, shifting $\bst$ and $\bar{\bst}$ 
as $\bst \to \bst-2[\lambda^{-1}]$ 
and $\bar{\bst} \to \bar{\bst}-2[\mu]$ 
in (\ref{BKP-diffFay1(iv)}) and 
multiply it by $e^{\xi(\bar{\bst},\mu)}
\tau(\bst,\bar{\bst}-2[\mu])/
\tau(\bst,\bar{\bst})$, we obtain 
the equation 
\begin{multline}
  (\bar{\rd}_1 + \bar{\rd}_1\log\Psi(\bst,\bar{\bst},\lambda))
  e^{-2D(\lambda)}\bar{\Psi}(\bst,\bar{\bst},\mu) \\
= (\bar{\rd}_1 - \bar{\rd}_1\log\Psi(\bst,\bar{\bst},\lambda))
  \bar{\Psi}(\bst,\bar{\bst},\mu). 
\label{BKP-auxlin-gen6}
\end{multline}
Expanded in powers of $\lambda$, 
these equations give rise to an infinite 
number of linear equations.  Among them, 
the equations from the $\lambda^{-1}$-terms 
give exactly the two-dimensional 
Schr\"odinger equation 
(\ref{BKP-auxlin2}): 
\beqnn
\begin{gathered}
  (\rd_1\bar{\rd}_1 + 2(\rd_1\bar{\rd}_1\log\tau))
  \Psi(\bst,\bar{\bst},\mu) = 0, \\
  (\rd_1\bar{\rd}_1 + 2(\rd_1\bar{\rd}_1\log\tau))
  \bar{\Psi}(\bst,\bar{\bst},\mu) = 0. 
\end{gathered}
\eeqnn

Thus the usual auxiliary linear equations 
(\ref{BKP-auxlin1}) and (\ref{BKP-auxlin2}) 
can be derived from the differential Fay identities.  
It should be noted that we have not 
fully used the generating functional 
equations (\ref{BKP-auxlin-gen1})
--(\ref{BKP-auxlin-gen4}), 
(\ref{BKP-auxlin-gen5}) and 
(\ref{BKP-auxlin-gen6}).  Namely, 
only even power terms of $\lambda$ 
in the first four equations 
and the $\lambda^{-1}$-terms in 
the last two equations are enough 
to recover all relevant auxiliary 
linear equations.   What meaning, 
then, do other terms have?  
The fact is that the other terms 
give no new equations.  For example, 
the equations for $n = 3$ in 
(\ref{BKP-auxlin-SS}) can be derived 
from the the corresponding equation 
for $n = 2$ by differentiating by 
$t_1$ and $\bar{t}_1$, respectively.

\subsection{Dispersionless Hirota equations 
and Hamilton-Jacobi equations}

The procedure of quasi-classical limit 
is a straightforward generalization 
of the case of the one-component 
BKP hierarchy \cite{Takasaki93}. 

As regards the Hirota formalism, 
we assume the quasi-classical ansatz 
\beq
  \tau_\hbar(\bst,\bar{\bst}) 
  = \exp\Bigl(\hbar^{-2}F(\bst,\bar{\bst}) 
    + O(\hbar^{-1})\Bigr) 
\label{BKP-tau-hbar}
\eeq
for the rescaled tau function 
$\tau_\hbar(\bst,\bar{\bst}) 
= \tau(\hbar,\hbar^{-1}\bst,\hbar^{-1}\bar{\bst})$. 
By this ansatz, we can derive the following 
dispersionless Hirota equations from 
the foregoing four differential Fay identities 
\cite{Takasaki06}: 
\begin{multline}
  \Bigl(\lambda + \mu - 2\rd_1(D(\lambda)+D(\mu))F\Bigr) 
  e^{4D(\lambda)D(\mu)F} \\
  = \frac{\lambda+\mu}{\lambda-\mu} 
    \Bigl(\lambda - \mu 
      - 2\rd_1(D(\lambda)-D(\mu))F\Bigr), 
\label{dBKP-Hirota1(i)}
\end{multline}
\begin{multline}
  \Bigl(\lambda + \mu 
    - 2\bar{\rd}_1(\bar{D}(\lambda)+\bar{D}(\mu))F\Bigr)
  e^{4\bar{D}(\lambda)\bar{D}(\mu)F} \\
  = \frac{\lambda+\mu}{\lambda-\mu} 
    \Bigl(\lambda - \mu 
      - 2\bar{\rd}_1(\bar{D}(\lambda)-\bar{D}(\mu))F\Bigr), 
\label{dBKP-Hirota1(ii)}
\end{multline}
\beq
  \Bigl(\lambda - 2\rd_1(D(\lambda)+\bar{D}(\mu))F\Bigr) 
  e^{4D(\lambda)\bar{D}(\mu)F} 
  = \lambda - 2\rd_1(D(\lambda)-\bar{D}(\mu))F, 
\label{dBKP-Hirota1(iii)}
\eeq
\beq
  \Bigl(\mu - 2\bar{\rd}_1(D(\lambda)+\bar{D}(\mu))F\Bigr) 
  e^{4D(\lambda)\bar{D}(\mu)F} 
  = \mu - 2\bar{\rd}_1(\bar{D}(\mu)-D(\lambda))F. 
\label{dBKP-Hirota1(iv)}
\eeq
By defining the $S$ functions 
$S(z) = S(\bst,\bar{\bst},z)$ and 
$\bar{S}(z) = \bar{S}(\bst,\bar{\bst},z)$ as 
\beqnn
  S(z) = \xi(\bst,z) - 2D(z)F, \quad 
  \bar{S}(z) = \xi(\bar{\bst},z) - 2\bar{D}(z)F, 
\eeqnn
(\ref{dBKP-Hirota1(i)}) -- (\ref{dBKP-Hirota1(iv)}) 
can be rewritten as 
\beq
  (\rd_1S(\lambda) + \rd_1S(\mu)) e^{4D(\lambda)D(\mu)F} 
  = \frac{\lambda+\mu}{\lambda-\mu} 
    (\rd_1S(\lambda) - \rd_1S(\mu)), 
\label{dBKP-Hirota2(i)}
\eeq
\beq
  (\bar{\rd}_1\bar{S}(\lambda) + \bar{\rd}_1\bar{S}(\mu) 
  e^{4\bar{D}(\lambda)\bar{D}(\mu)F} 
  = \frac{\lambda+\mu}{\lambda-\mu} 
    (\bar{\rd}_1\bar{S}(\lambda) - \bar{\rd}_1\bar{S}(\mu)), 
\label{dBKP-Hirota2(ii)}
\eeq
\beq
  (\rd_1S(\lambda) + \rd_1\bar{S}(\mu)) 
  e^{4D(\lambda)\bar{D}(\mu)F} 
  = \rd_1S(\lambda) - \rd_1\bar{S}(\mu), 
\label{dBKP-Hirota2(iii)}
\eeq
\beq
  (\bar{\rd}_1\bar{S}(\mu) + \bar{\rd}_1S(\lambda)) 
  e^{4D(\lambda)\bar{D}(\mu)F} 
  = \bar{\rd}_1\bar{S}(\mu) - \bar{\rd}_1S(\lambda). 
\label{dBKP-Hirota2(iv)}
\eeq
Actually, (\ref{dBKP-Hirota2(i)}) was first 
obtained by Bogdanov and Konopelchenko 
\cite{BK05} by the $\bar{\rd}$-dressing method, 
and rederived, along with the other equations, 
by Takasaki \cite{Takasaki06} by the method 
presented here.  Later on, Chen and Tu \cite{CT06} 
studied these equations by the kernel method 
of Carroll and Kodama \cite{CK95}. 

Quasi-classical limit is also achieved 
in the Lax formalism.  The ansatz 
for the wave functions again takes the WKB form 
\beq
\begin{gathered}
  \Psi_\hbar(\bst,\bar{\bst},z) 
  = \exp\Bigl(\hbar^{-1}S(z) + O(\hbar^0)\Bigr), \\
  \bar{\Psi}_\hbar(\bst,\bar{\bst},z) 
  = \exp\Bigl(\hbar^{-1}\bar{S}(z) + O(\hbar^0)\Bigr) 
\end{gathered}
\label{BKP-hbar-Psi}
\eeq
for the rescaled wave functions 
$\Psi_\hbar(\bst,\bar{\bst},z) 
= \Psi(\hbar,\hbar^{-1}\bst,\hbar^{-1}\bar{\bst},z)$ 
and $\bar{\Psi}_\hbar(\bst,\bar{\bst},z) 
= \bar{\Psi}(\hbar,\hbar^{-1}\bst,\hbar^{-1}\bar{\bst},z)$. 
The foregoing $S$ functions $S = S(z),\bar{S}(z)$ 
show up here as the phase functions of 
the WKB approximation, and satisfy 
the Hamilton-Jacobi equations 
\beq
\begin{gathered}
  \rd_{2n+1}S = \calB_{2n+1}(\rd_1S), \quad 
  \bar{\rd}_{2n+1}S = \bar{\calB}_{2n+1}(\bar{\rd}_1S),\\
  \rd_1\bar{\rd}_1S = u 
\end{gathered}
\eeq
with the classical Hamiltonians 
\beqnn
  \calB_{2n+1}(p) 
  = \lim_{\hbar\to 0}B_{\hbar,2n+1}(p), \quad 
  \bar{\calB}_{2n+1}(\bar{p}) 
  = \lim_{\hbar\to 0}\bar{B}_{\hbar,2n+1}(\bar{p}) 
\eeqnn
obtained from the (rescaled) operators 
$B_{\hbar,2n+1} = B_{\hbar,2n+1}(\hbar\rd_1)$ and 
$\bar{B}_{\hbar,2n+1} = \bar{B}_{\hbar,2n+1}(\hbar\bar{\rd}_1)$. 
Hamilton-Jacobi equations of this type 
were also studied by Konopelchenko and 
Moro \cite{KM04} in the context of nonlinear optics.  

Though the structure of the dispersionless 
Hirota equations look considerably 
different from those of the dispersionless 
KP and Toda hierarchies, one can anyhow 
show that they are, in fact, equivalent 
to the Hamilton-Jacobi equations 
\cite{Takasaki06}.

\section{DKP hierarchy and differential Fay identities}

\subsection{Bilinear equations for tau function}

Our final case study is focused on the DKP hierarchy. 
This is a subsystem of one of Jimbo and Miwa's 
integrable hierarchies with with $D_\infty$ symmetries 
\cite{JM83}.  The master system proposed 
by Jimbo and Miwa is formulated in terms of 
tau functions $\tau(s,\bst)$ with a discrete variable 
$s \in \ZZ$ and a set of continuous variables 
$\bst = (t_1,t_2,\ldots)$.  
These tau functions satisfy an infinite number 
of Hirota equations (or an equivalent set of 
bilinear equations of the contour integral type), 
which can be divided to three subsets, namely, 
the equations for $\tau(2s,\bst)$'s, 
the equations for $\tau(2s+1,\bst)$'s 
and the equations of the coupled type inbetween. 
The Hirota equations of the first and second types 
are actually identical;  it is this system of 
Hirota equations that we call ``the DKP hierarchy''.  
Thus Jimbo and Miwa's master system contains 
two copies of the DKP hierarchy as subsystems, 
and may be called a two-component DKP hierarchy.  

Since being a relatively less known member 
of the big family of integrable hierarchies, 
the DKP hierarchy has been rediscovered 
with a different name such as 
``the coupled KP hierarchy'' 
\cite{HO91,Kakei99,Kakei00,IWS02-03} 
and ``the Pfaff lattice''\cite{AHvM99,ASvM99}. 
Kac and van de Leur \cite{KvdL98} called 
Jimbo and Miwa's master system 
``the charged BKP hierarchy'', 
and formulated a multi-component hierarchy 
in which the BKP, DKP and charged BKP hierarchies 
and their multi-component analogues are unified. 

Let $\tau(s,\bst)$ denote the tau function 
of the DKP hierarchy.  Note that it amounts 
to $\tau(2s,\bst)$ or $\tau(2s+1,\bst)$ 
in Jimbo and Miwa's master system.  
This tau function satisfies the bilinear equation 
\begin{multline}
  \oint\frac{dz}{2\pi i}
  z^{2s'-2s-2}e^{\xi(\bst'-\bst,z)} 
  \tau(s'-1,\bst'-[z^{-1}])\tau(s,\bst+[z^{-1}]) \\
\mbox{} 
+ \oint\frac{dz}{2\pi i}
  z^{2s-2s'-2}e^{\xi(\bst-\bst',z)}
  \tau(s',\bst'+[z^{-1}])\tau(s-1,\bst-[z^{-1}]) 
= 0. 
\label{DKP-bilin-tau}
\end{multline}
Here and in the following, the notations 
are mostly the same as used for the KP hierarchy.   
The integrals on the left hand side are contour integrals 
along a sufficiently large circle $|z| = R$ 
(or a formal algebraic operator 
extracting the coefficient of $\lambda^{-1}$).  
As in the other cases, this bilinear equation, 
too, can be converted to the Hirota form 
\begin{multline}
  \sum_{n=0}^\infty h_n(-2\bsa) 
  h_{n+2s'-2s-1}(\tilde{D}_{\bst})e^{\langle\bsa,D_{\bst}\rangle} 
  \tau(s,\bst)\cdot\tau(s'-1,\bst) \\
\mbox{}
+ \sum_{n=0}^\infty h_n(2\bsa) 
  h_{n+2s-2s'-1}(-\tilde{D}_{\bst})e^{\langle\bsa,D_{\bst}\rangle}
  \tau(s-1,\bst)\cdot\tau(s',\bst) 
= 0. 
\label{DKP-Hirota-gen}
\end{multline}
For example, if $s' = s+1$, terms linear in $a_n$ 
give the special Hirota equations 
\begin{multline}
  (D_1D_n - 2h_{n+1}(\tilde{D}_{\bst}))
  \tau(s,\bst)\cdot\tau(s,\bst) \\
  - 2h_{n-3}(-\tilde{D}_{\bst})
    \tau(s-1,\bst)\cdot\tau(s+1,\bst) 
  = 0 
\label{DKP-Hirota-special}
\end{multline}
for $n = 3,4,\ldots$.  

Note that the left hand side 
of (\ref{DKP-Hirota-special}) coincide 
with that of the special Hirota equations 
(\ref{KP-Hirota-special}) of the KP hierarchy.  
In this respect, the right hand side 
may be thought of as a coupling term 
among the tau functions $\tau(s,\bst)$ 
with different indices;  the name 
``coupled KP hierarchy'' originates 
in such an interpretation.  
This is, however, a rather superficial analogy.  
Both systems are considerably different. 
For example, as the name ``Pfaff lattice'' 
indicates, its relevant equations and 
solutions exhibit a Pfaffian structure 
\cite{HO91,Kakei99,Kakei00,AHvM99,ASvM99,
vdL01,AKvM02,KM06,KP07}
unlike the determinantal structure of 
the KP hierarchy.

Let us mention that the BKP hierarchy 
is also known to have a Pfaffian structure. 
The difference between the BKP and DKP hierarchies 
stems from the type of free fermions behind 
the tau functions \cite{JM83,KvdL98}.  
Namely, whereas the BKP hierarchy is related 
to neutral fermions, the DKP hierarchy 
is formulated in terms of charged fermions. 
The charged fermion system is also used 
for the KP hierarchy; this explains 
the superficial similarity between 
the KP and DKP hierarchies. Actually, 
Jimbo and Miwa \cite{JM83} used 
three types of free fermion systems 
(charged fermions, neutral fermions and 
two-component charged fermions) to 
formulate three integrable hierarchies with 
$D_\infty$ symmetries.  The DKP hierarchy 
is embedded in their first hierarchy 
formulated by charged fermions.  
It is interesting that their second hierarchy,  
formulated by neutral fermions, contains 
two copies of the two-component BKP hierarchy.

\subsection{Differential Fay identities} 

Two differential Fay identities are 
obtained from the bilinear equation 
(\ref{DKP-bilin-tau}) by the following procedure: 
\begin{itemize}
\item[(i)] Differentiate the bilinear equation 
by $t'_1$ and specialize $s',\bst'$ as 
$s' = s+1$ and 
$\bst' = \bst + [\lambda^{-1}] + [\mu^{-1}]$. 
\item[(ii)] Differentiate the bilinear equation 
by $t'_1$ and specialize $s',\bst'$ as 
$s' = s$ and 
$\bst' = \bst + [\lambda^{-1}] + [\mu^{-1}]$. 
\end{itemize}

By (i), the bilinear equation reduces to 
\begin{multline*}
  \oint\frac{dz}{2\pi i}
  \frac{\lambda\mu}{(z-\lambda)(z-\mu)}
  \Bigl(z\tau(s,\bst+[\lambda^{-1}]+[\mu^{-1}]-[z^{-1}]) \\
   + (\rd_1\tau)(s,\bst+[\lambda^{-1}]+[\mu^{-1}]-[z^{-1}])
  \Bigr)\tau(s,\bst+[z^{-1}]) \\
\mbox{} 
+ \oint\frac{dz}{2\pi i}
  \frac{(z-\lambda)(z-\mu)}{\lambda\mu z^4}
  \Bigl(-z\tau(s+1,\bst+[\lambda^{-1}]+[\mu^{-1}]+[z^{-1}])\\
    + (\rd_1\tau)(s+1,\bst+[\lambda^{-1}]+[\mu^{-1}]+[z^{-1}])
  \Bigr)\tau(s-1,\bst-[z^{-1}]) 
= 0. 
\end{multline*}
Residue calculus and some algebra yield 
the equation 
\begin{multline}
  \frac{\tau(s,\bst+[\lambda^{-1}]+[\mu^{-1}])\tau(s,\bst)}
       {\tau(s,\bst+[\lambda^{-1}])\tau(s,\bst+[\mu^{-1}])} \\
  - \frac{1}{\lambda^2\mu^2} 
    \frac{\tau(s+1,\bst+[\lambda^{-1}]+[\mu^{-1}])\tau(s-1,\bst)}
         {\tau(s,\bst+[\lambda^{-1}])\tau(s,\bst+[\mu^{-1}])} \\
  = 1 
    - \frac{1}{\lambda-\mu}\rd_1\log
      \frac{\tau(s,\bst+[\lambda^{-1}])}
           {\tau(s,\bst+[\mu^{-1}])}. 
\label{DKP-diffFay1(i)}
\end{multline}
In the same way, (ii) leads to the equation 
\begin{multline*}
  \frac{\lambda^2}{\lambda-\mu}
  \frac{\tau(s,\bst)\tau(s-1,\bst+[\lambda^{-1}]-[\mu^{-1}])}
       {\tau(s,\bst+[\lambda^{-1}])\tau(s-1,\bst-[\mu^{-1}])}\\
- \frac{\mu^2}{\lambda-\mu}
  \frac{\tau(s,\bst+[\lambda^{-1}]-[\mu^{-1}])\tau(s-1,\bst)}
       {\tau(s,\bst+[\lambda^{-1}])\tau(s-1,\bst-[\mu^{-1}])} \\
= \lambda + \mu 
  - \rd_1\log
    \frac{\tau(s,\bst+[\lambda^{-1}])}{\tau(s-1,\bst-[\mu^{-1}])}, 
\end{multline*}
but from the point of view of symmetry, 
it will be better to shift $\bst$ as 
$\bst \to \bst + [\mu^{-1}]$.  
This gives the equation 
\begin{multline}
  \frac{\lambda^2}{\lambda-\mu}
  \frac{\tau(s,\bst+[\mu^{-1}])\tau(s-1,\bst+[\lambda^{-1}])}
       {\tau(s,\bst+[\lambda^{-1}]+[\mu^{-1}])\tau(s-1,\bst)} \\
- \frac{\mu^2}{\lambda-\mu}
  \frac{\tau(s,\bst+[\lambda^{-1}])\tau(s-1,\bst+[\mu^{-1}])}
       {\tau(s,\bst+[\lambda^{-1}]+[\mu^{-1}])\tau(s-1,\bst)} \\
= \lambda + \mu 
  - \rd_1\log
    \frac{\tau(s,\bst+[\lambda^{-1}]+[\mu^{-1}])}{\tau(s-1,\bst)}.
\label{DKP-diffFay1(ii)}
\end{multline}

(\ref{DKP-diffFay1(i)}) was first derived 
by Adler, Horozov and van Moerbeke 
in their study of the Pfaff lattice \cite{AHvM99} . 
As they pointed out, this equation resembles 
the differential Fay identity (\ref{KP-diffFay1}) 
of the KP hierarchy; this is another aspect of 
the aforementioned similarity with the KP hierarchy.  
The only difference is the presence of 
the second term on the left hand side. 
This is, however, an essential difference.  

(\ref{DKP-diffFay1(ii)}) has no analogue 
in the KP hierarchy, etc.  A remarkable 
characteristic of this equation and 
(\ref{DKP-diffFay1(i)}) is that 
there are two terms, rather than just one, 
that contain the double shift 
$\bst+[\lambda^{-1}]+[\mu^{-1}]$ 
of $\bst$.  This apparently small difference 
eventually affects every aspect of 
this system and its dispersionless analogue.  

Lastly, using the operator $D(z)$, 
we can rewrite these differential 
Fay identities as follows: 
\begin{multline}
  \exp\Bigl((e^{D(\lambda)}-1)(e^{D(\mu)}-1)\log\tau\Bigr) \\
  - \frac{1}{\lambda^2\mu^2}\exp\Bigl(
     (e^{D(\lambda)+\rd_s}-1)(e^{D(\mu)+\rd_s}-1) 
     e^{-\rd_s}\log\tau\Bigr) \\
  = 1 
    - \frac{\rd_1(e^{D(\lambda)}-e^{D(\mu)})\log\tau}{\lambda-\mu}, 
\label{DKP-diffFay2(i)}
\end{multline}
\begin{multline}
  \frac{\lambda^2}{\lambda-\mu} 
  \exp\Bigl(-(e^{D(\lambda)}-1)(e^{D(\mu)}-e^{-\rd_s})\log\tau\Bigr) \\
  - \frac{\mu^2}{\lambda-\mu} 
    \exp\Bigl(-(e^{D(\lambda)}-e^{-\rd_s})(e^{D(\mu)}-1)\log\tau\Bigr)\\
  = \lambda + \mu 
    - \rd_1(e^{D(\lambda)+D(\mu)}-e^{-\rd_s})\log\tau. 
\label{DKP-diffFay2(ii)}
\end{multline}

\subsection{Auxiliary linear equations}

The rather complicated structure 
of the differential Fay identities 
is closely related to the fact 
that an underlying auxiliary 
linear problem is vector-valued. 
This auxiliary linear problem 
is substantially the same as the one 
first discovered by Kakei \cite{Kakei00} 
in an inverse scattering formalism, 
and just a special case of the work 
of kac and van de Leur \cite{KvdL98} 
on multi-component DKP/BKP hierarchies.    

We introduce the wave functions
\beqnn
\begin{gathered}
  \Psi_1(s,\bst,z) 
  = \frac{\tau(s,\bst-[z^{-1}])}{\tau(s,\bst)}
    z^{2s}e^{\xi(\bst,z)}, \\
  \Psi_2(s,\bst,z) 
  = \frac{\tau(s-1,\bst-[z^{-1}])}{\tau(s,\bst)} 
    z^{2s-2}e^{\xi(\bst,z)} 
\end{gathered}
\eeqnn
and the duals 
\beqnn
\begin{gathered}
  \Psi_1^*(s,\bst,z) 
  = \frac{\tau(s+1,\bst+[z^{-1}])}{\tau(s,\bst)} 
    z^{-2s-2}e^{-\xi(\bst,z)}, \\
  \Psi_2^*(s,\bst,z) 
  = \frac{\tau(s-1,\bst+[z^{-1}])}{\tau(s,\bst)} 
    z^{-2s}e^{-\xi(\bst,z)}. 
\end{gathered}
\eeqnn
By suitably shifting $s'$ and $s$, 
the bilinear equation (\ref{DKP-bilin-tau}) 
for the tau function yields 
the bilinear equations 
\beqnn
  \oint\frac{dz}{2\pi i}
  \Psi_1(s',\bst',z)\Psi_1^*(s,\bst,z) 
+ \oint\frac{dz}{2\pi i}
  \Psi_1^*(s',\bst,z)\Psi_1(s,\bst,z) = 0, \\
  \oint\frac{dz}{2\pi i}
  \Psi_1(s',\bst',z)\Psi_2^*(s,\bst,z) 
+ \oint\frac{dz}{2\pi i}
  \Psi_1^*(s',\bst',z)\Psi_2(s,\bst,z) = 0, \\
  \oint\frac{dz}{2\pi i}
  \Psi_2(s',\bst',z)\Psi_1^*(s,\bst,z) 
+ \oint\frac{dz}{2\pi i}
  \Psi_2^*(s',\bst',z)\Psi_1(s,\bst,z) = 0, \\
  \oint\frac{dz}{2\pi i}
  \Psi_2(s',\bst',z)\Psi_2^*(s,\bst,z) 
+ \oint\frac{dz}{2\pi i}
  \Psi_2^*(s',\bst',z)\Psi_2(s,\bst,z) = 0 
\eeqnn
for these wave functions.  Moreover, 
these four equations can be cast 
into the matrix form 
\begin{multline}
  \oint\frac{dz}{2\pi i}
  \left(\begin{array}{cc}
  \Psi_1(s',\bst',z) & \Psi_1^*(s',\bst',z)\\
  \Psi_2(s',\bst',z) & \Psi_2^*(s',\bst',z)
  \end{array}\right) 
  \left(\begin{array}{cc}
  \Psi_1^*(s,\bst,z) & \Psi_2^*(s,\bst,z)\\
  \Psi_1(s,\bst,z) & \Psi_2(s,\bst,z)
  \end{array}\right) 
= 0. 
\label{DKP-bilin-Psi}
\end{multline}
This matrix bilinear equation and 
its building blocks are reminiscent 
of those of the charged two-component 
KP hierarchy \cite{DJKM-KP}.  
There is, however, an unnegligible difference.  
In the case of the two-component KP hierarchy 
with charge $(2s,-2s)$, 
the $(2,1)$ and $(1,2)$ elements of 
the matrix wave functions should have 
$z^{2s-1}$ and $z^{-2s-1}$ rather than 
$z^{2s-2}$ and $z^{-2s-2}$ in the foregoing 
definition of the wave functions.  
Therefore if one interprets this system 
as a reduction of the two-component 
KP hierarchy, one should remember that 
this is a very special reduction.  
Actually, Adler, Horozov and van Moerbeke 
\cite{AHvM99} obtained the Pfaff lattice 
in a ``deep stratum'' of the phase space 
of the Toda hierarchy.  

As Kac and van de Leur did 
in a more general setting \cite{KvdL98}, 
one can derive auxiliary linear equations 
of the form 
\begin{multline}
  \rd_n
  \left(\begin{array}{cc}
  \Psi_1(s,\bst,z) & \Psi_1^*(s,\bst,z)\\
  \Psi_2(s,\bst,z) & \Psi_2^*(s,\bst,z)
  \end{array}\right) \\
= \left(\begin{array}{cc}
  A_n & B_n \\
  C_n & D_n 
  \end{array}\right)
  \left(\begin{array}{cc}
  \Psi_1(s,\bst,z) & \Psi_1^*(s,\bst,z)\\
  \Psi_2(s,\bst,z) & \Psi_2^*(s,\bst,z)
  \end{array}\right), 
\label{DKP-auxlin}
\end{multline}
where $A_n = A_n(s,\bst,\rd_1),
B_n = B_n(s,\bst,\rd_1),
C_n = C_n(s,\bst,\rd_1),
D_n = D_n(s,\bst,\rd_1)$  are 
differential operator with respect to $t_1$.  
One can specify the structure of 
these operators in more detail 
by introducing dressing operators 
for the wave functions.  In particular, 
as Kakei observed \cite{Kakei00}, 
$A_n,B_n,C_n,D_n$ turn out to be operators 
of the form 
\beq
\begin{aligned}
  &A_n = \rd_1^n + O(\rd_1^{n-2}), &\quad 
  &B_n = O(\rd_1^{n-2}), \\
  &C_n = O(\rd_1^{n-2}), &\quad 
  &D_n = - (-\rd_1)^n + O(\rd_1^{n-2}) 
\end{aligned}
\label{DKP-ABCD-order}
\eeq
and satisfy the algebraic relations 
\beq
\begin{aligned}
  &A_n^* = - D_n, &\quad &D_n^* = - A_n, \\
  &B_n^* = B_n,   &\quad &C_n^* = C_n, 
\end{aligned}
\label{DKP-ABCD-relation}
\eeq
where $A_n^*,B_n^*,C_n^*,D_n^*$ denote 
the formal adjoint of $A_n,B_n,C_n,D_n$.

\subsection{Differential Fay identities and 
auxiliary linear equations}

We now translate the differential 
Fay identities (\ref{DKP-diffFay1(i)}) 
and (\ref{DKP-diffFay1(ii)}) 
to the language of the wave functions.  
As it turns out, each of the differential 
Fay identities can be converted to 
two different equations (hence altgether 
four equations) for the wave functions.  
These equations give a generating 
functional expression of the auxiliary 
linear equations presented above.  

Let us first consider (\ref{DKP-diffFay1(i)}).  
Shifting $\bst$ as 
$\bst \to \bst-[\lambda^{-1}]-[\mu^{-1}]$, 
we have the equation 
\begin{multline*}
  \frac{\tau(s,\bst-[\lambda^{-1}]-[\mu^{-1}])\tau(s,\bst)}
       {\tau(s,\bst-[\lambda^{-1}])\tau(s,\bst-[\mu^{-1}])}\\
  - \frac{1}{\lambda^2\mu^2} 
    \frac{\tau(s+1,\bst)\tau(s-1,\bst-[\lambda^{-1}]-[\mu^{-1}])}
         {\tau(s,\bst-[\lambda^{-1}])\tau(s,\bst-[\mu^{-1}])}\\
  = 1 
    + \frac{1}{\lambda-\mu}\rd_1\log
      \frac{\tau(s,\bst-[\lambda^{-1}])}
           {\tau(s,\bst-[\mu^{-1}])}, 
\end{multline*}
and multiplying this equation by 
$(\lambda-\mu)\mu^{2s}e^{\xi(\bst,\mu)}
\tau(s,\bst-[\mu^{-1}])/\tau(s,\bst)$, 
we obtain the equation 
\begin{multline}
  \lambda e^{-D(\lambda)}\Psi_1(s,\bst,\mu) 
  - \lambda^{-1}\frac{\tau(s+1,\bst)}{\tau(s,\bst)} 
    e^{-D(\lambda)}\Psi_2(s,\bst,\mu) \\
  = (\rd_1\log\Psi_1(s,\bst,\lambda) - \rd_1)
    \Psi_1(s,\bst,\mu). 
\label{DKP-auxlin-gen1}
\end{multline}
Similarly, multiplying both hand side 
of (\ref{DKP-diffFay1(i)}) (without 
shifting $\bst$) by 
$(\lambda-\mu)\mu^{-2s}e^{-\xi(\bst,\mu)}
\tau(s,\bst+[\mu^{-1}])/\tau(s,\bst)$ 
yields the equation 
\begin{multline}
  \lambda e^{D(\lambda)}\Psi^*_2(s,\bst,\mu) 
  - \lambda^{-1}\frac{\tau(s-1,\bst)}{\tau(s,\bst)}
    e^{D(\lambda)}\Psi^*_1(s,\bst,\mu) \\
  = (\rd_1 - \rd_1\log\Psi^*_2(s,\bst,\lambda))
    \Psi^*_2(s,\bst,\mu). 
\label{DKP-auxlin-gen2}
\end{multline}
Note that, unlike the previous cases, 
there are two terms containing 
$e^{\pm D(\lambda)}$.  They stem from 
the two terms in (\ref{DKP-diffFay1(i)}) 
containing the double shift 
$\bst+[\lambda^{-1}]+[\mu^{-1}]$ of $\bst$.  

We now consider (\ref{DKP-diffFay1(ii)}) 
or, rather its original form 
\begin{multline*}
  \frac{\lambda^2}{\lambda-\mu}
  \frac{\tau(s,\bst)\tau(s-1,\bst+[\lambda^{-1}]-[\mu^{-1}])}
       {\tau(s,\bst+[\lambda^{-1}])\tau(s-1,\bst-[\mu^{-1}])}\\
- \frac{\mu^2}{\lambda-\mu}
  \frac{\tau(s,\bst+[\lambda^{-1}]-[\mu^{-1}])\tau(s-1,\bst)}
       {\tau(s,\bst+[\lambda^{-1}])\tau(s-1,\bst-[\mu^{-1}])}\\
= \lambda + \mu 
  - \rd_1\log
    \frac{\tau(s,\bst+[\lambda^{-1}])}{\tau(s-1,\bst-[\mu^{-1}])}. 
\end{multline*}
Multiplying this equation by 
$\lambda^{-2s}e^{-\xi(\bst,\lambda)}
\tau(s,\bst+[\lambda^{-1}])/\tau(s-1,\bst)$ 
yields the equation 
\begin{multline*}
  \mu^{-1}\frac{\tau(s,\bst)}{\tau(s-1,\bst)}
  e^{-D(\mu)}\Psi^*_2(s,\bst,\lambda) 
  - \mu e^{-D(\mu)}\Psi^*_1(s-1,\bst,\lambda) \\
  = (\rd_1 - \rd_1\log\Psi_1(s-1,\bst,\mu))
    \Psi^*_1(s-1,\bst,\lambda). 
\end{multline*}
Exchanging $\lambda$ and $\mu$ and 
shifting $s$ as $s \to s+1$, 
we eventually obtain the equation 
\begin{multline}
  \lambda e^{-D(\lambda)}\Psi^*_1(s,\bst,\mu) 
  - \lambda^{-1}\frac{\tau(s+1,\bst)}{\tau(s,\bst)}
   e^{-D(\lambda)}\Psi^*_2(s,\bst,\mu) \\
  = (\rd_1\log\Psi_1(s,\bst,\lambda) - \rd_1)
    \Psi^*_1(s,\bst,\mu), 
\label{DKP-auxlin-gen3}
\end{multline}
which has the same structure as 
(\ref{DKP-auxlin-gen1}).  
Similarly, multiplying the same equation 
as above by 
$\mu^{2s-2}e^{\xi(\bst,\mu)}
\tau(s-1,\bst-[\mu^{-1}])/\tau(s,\bst)$, 
we obtain the equation 
\begin{multline}
  \lambda e^{D(\lambda)}\Psi_2(s,\bst,\mu) 
  - \lambda^{-1}\frac{\tau(s-1,\bst)}{\tau(s,\bst)}
    e^{D(\lambda)}\Psi_1(s,\bst,\mu) \\
  = (\rd_1 - \rd_1\log\Psi^*_2(s,\bst,\lambda))
    \Psi_2(s,\bst,\mu), 
\label{DKP-auxlin-gen4}
\end{multline}
which can be compared with 
(\ref{DKP-auxlin-gen2}).  

These results (\ref{DKP-auxlin-gen1})
--(\ref{DKP-auxlin-gen4}) show that 
the vector-valued wave functions 
\beqnn
  \left(\begin{array}{c}
  \Phi_1 \\
  \Phi_2
  \end{array}\right)
  = \left(\begin{array}{c}
    \Psi_1(s,\bst,\mu)\\
    \Psi_2(s,\bst,\mu)
    \end{array}\right),\, 
    \left(\begin{array}{c}
    \Psi^*_1(s,\bst,\mu)\\
    \Psi^*_2(s,\bst,\mu)
    \end{array}\right) 
\eeqnn
satisfy the same linear equations 
\beqnn
\begin{gathered}
  \lambda e^{-D(\lambda)}\Phi_1 
  - \lambda^{-1}\frac{\tau(s+1,\bst)}{\tau(s,\bst)} 
    e^{-D(\lambda)}\Phi_2 
  = (\rd_1\log\Psi_1(s,\bst,\lambda) - \rd_1)\Phi_1, \\
  \lambda e^{D(\lambda)}\Phi_2 
  - \lambda^{-1}\frac{\tau(s-1,\bst)}{\tau(s,\bst)}
    e^{D(\lambda)}\Phi_1 
  = (\rd_1 - \rd_1\log\Psi^*_2(s,\bst,\lambda))\Phi_2. 
\end{gathered}
\eeqnn
Expanded in powers of $\lambda$, 
these equations give rise to 
an infinite number of linear equations 
of the form 
\beq
\begin{gathered}
  h_n(-\tilde{\rd}_{\bst})\Phi_1 
  - \frac{\tau(s+1,\bst)}{\tau(s,\bst)}
    h_{n-2}(-\tilde{\rd}_{\bst})\Phi_2 = v_n\Phi_1, \\
  h_n(\tilde{\rd}_{\bst})\Phi_2 
  - \frac{\tau(s-1,\bst)}{\tau(s,\bst)}
    h_{n-2}(\tilde{\rd}_{\bst})\Phi_1 = u_n\Phi_2 
\end{gathered}
\label{DKP-auxlin-SS}
\eeq
for $n = 2,3,\ldots$, 
where $v_n$ and $u_n$ are defined by 
\beqnn
  v_n = \rd_1h_{n-2}(-\tilde{\rd}_{\bst})\log\tau(s,\bst), 
  \quad 
  u_n = \rd_1h_{n-2}(\tilde{\rd}_{\bst})\log\tau(s,\bst). 
\eeqnn
The lowest equations (for $n = 2$) read
\beqnn
\begin{gathered}
  \frac{1}{2}(\rd_1^2 - \rd_2)\Phi_1 
  - \frac{\tau(s+1,\bst)}{\tau(s,\bst)}\Phi_2 
  = - (\rd_1^2\log\tau(s,\bst))\Phi_1, \\
  \frac{1}{2}(\rd_1^2 + \rd_2)\Phi_2 
  - \frac{\tau(s-1,\bst)}{\tau(s,\bst)}\Phi_1 
  = - (\rd_1^2\log\tau(s,\bst))\Phi_2, 
\end{gathered}
\eeqnn
which one can rewrite as 
\beqnn
  \rd_2\left(\begin{array}{c}
  \Phi_1\\
  \Phi_2
  \end{array}\right) 
 = \left(\begin{array}{cc}
   \rd_1^2 + 2\rd_1^2\log\tau(s,\bst) & 
     -2 \frac{\tau(s+1,\bst)}{\tau(s,\bst)}\\
   2 \frac{\tau(s-1,\bst)}{\tau(s,\bst)} &
     \rd_1^2 - 2\rd_1^2\log\tau(s,\bst) 
   \end{array}\right)
   \left(\begin{array}{c}
   \Phi_1\\
   \Phi_2
   \end{array}\right) . 
\eeqnn
This is exactly the lowest equation of 
the auxiliary linear problem  (\ref{DKP-auxlin}).  
Higher equations can be derived recursively 
by the same procedure as in the case of 
the KP hierarchy etc.

\subsection{Quasi-classical limit --- encounter with difficulty}

Quasi-classical limit can be achieved 
by assuming the ansatz 
\beq
  \tau_\hbar(s,\bst) 
  = \exp\Bigl(\hbar^{-2}F(s,\bst) + O(\hbar^{-1})\Bigr) 
\label{DKP-tau-hbar}
\eeq
for the rescaled tau function $\tau_\hbar(s,\bst) 
= \tau(\hbar,\hbar^{-1}s,\hbar^{-1}\bst)$.  
The differential Fay identities, 
written in the form of (\ref{DKP-diffFay2(i)}) 
and (\ref{DKP-diffFay2(ii)}), yield 
the dispersionless Hirota equations 
\beq
  e^{D(\lambda)D(\mu)F} 
  - \lambda^{-2}\mu^{-2} 
    e^{(D(\lambda)+\rd_s)(D(\mu)+\rd_s)F} 
  = 1 
    - \frac{\rd_1(D(\lambda)-D(\mu))F}{\lambda-\mu}, 
\label{dDKP-Hirota(i)}
\eeq
\begin{multline}
  \frac{\lambda^2}{\lambda-\mu}e^{-D(\lambda)(D(\mu)+\rd_s)F}
  - \frac{\mu^2}{\lambda-\mu}e^{-(D(\lambda)+\rd_s)D(\mu)F} \\
  = \lambda + \mu 
    - \rd_1(D(\lambda)+D(\mu)+\rd_s)F 
\label{dDKP-Hirota(ii)}
\end{multline}
for the $F$ function.  
The first equation obviously resembles 
its counterpart (\ref{dKP-Hirota}) 
for the KP hierarchy; 
the only difference is the presence of 
the second term on the left hand side. 
On the other hand, 
the second equation seems to have 
no analogue in the previous cases.  

Let us examine these equations 
in  more detail.  It is convenient 
(and natural) to introduce 
the $S$ function 
\beqnn
  S(z) = \xi(\bst,z) + 2s\log z - D(z)F 
\eeqnn
and rewrite the two equations as 
\beq
  e^{D(\lambda)D(\mu)F} 
  (1 - e^{\rd_s^2F-\rd_sS(\lambda)-\rd_sS(\mu)})
  = \frac{\rd_1S(\lambda)-\rd_1S(\mu)}{\lambda-\mu},
\label{dDKP-Hirota(i)-S} 
\eeq
\beq
  \frac{e^{-D(\lambda)D(\mu)F}}{\lambda-\mu}
  (e^{\rd_sS(\lambda)}-e^{\rd_sS(\mu)})
  = \rd_1S(\lambda) + \rd_1S(\mu) - \rd_s\rd_1F. 
\label{dDKP-Hirota(ii)-S}
\eeq
One can eliminate $e^{D(\lambda)D(\mu)F}$ 
by multiplying both hand sides of these equations. 
This yields the relation 
\begin{multline*}
  (1 - e^{\rd_s^2F-\rd_sS(\lambda)-\rd_sS(\mu)})
  (e^{\rd_sS(\lambda)}-e^{\rd_sS(\mu)}) \\
  = (\rd_1S(\lambda)-\rd_1S(\mu))
    (\rd_1S(\lambda) + \rd_1S(\mu) - \rd_s\rd_1F), 
\end{multline*}
which one can rewrite as 
\begin{multline*}
  (\rd_1S(\lambda))^2 - (\rd_1\rd_sF)(\rd_1S(\lambda)) 
  - e^{\rd_sS(\lambda)} - e^{\rd_s^2F-\rd_sS(\lambda)} \\
= (\rd_1S(\mu))^2 - (\rd_1\rd_sF)(\rd_1S(\mu)) 
  - e^{\rd_sS(\mu)} - e^{\rd_s^2F-\rd_sS(\mu)}. 
\end{multline*}
Since $\lambda$ and $\mu$ are separated 
to the left hand side and the right hand side, 
both hand sides of this equation are 
actually independent of $\lambda$ and $\mu$. 
One can determine their value by letting 
$\lambda,\mu \to \infty$.  We thus obtain 
the equation 
\begin{multline}
  (\rd_1S(\mu))^2 - (\rd_1\rd_sF)(\rd_1S(\mu)) 
  - e^{\rd_sS(\mu)} - e^{\rd_s^2F-\rd_sS(\mu)} \\
  = - 2\rd_1^2F + \frac{1}{2}\rd_2\rd_sF 
    - \frac{1}{2}(\rd_1\rd_sF)^2 
\label{dDKP-Hirota(iii)-S}
\end{multline}
as a consequence of (\ref{dDKP-Hirota(i)-S}) 
and (\ref{dDKP-Hirota(ii)-S}).  
Actually, this procedure can be reversed.  
Namely, assuming that 
(\ref{dDKP-Hirota(i)-S}) holds, 
one can easily recover 
(\ref{dDKP-Hirota(ii)-S}) from  
(\ref{dDKP-Hirota(iii)-S}).  
Thus under (\ref{dDKP-Hirota(i)-S}), 
(\ref{dDKP-Hirota(ii)-S}) and 
(\ref{dDKP-Hirota(iii)-S}) are equivalent.  

(\ref{dDKP-Hirota(iii)-S}) may be 
thought of as a quadratic constraint 
to (\ref{dDKP-Hirota(i)-S}).  
This constraint can be solved for 
either $\rd_1S(\mu)$ or $e^{\rd_sS(\mu)}$, 
which becomes a non-polynomial function 
of the other.  For example, solving it 
for $e^{\rd_sS(\mu)}$ gives 
\beq
  e^{\rd_sS(\mu)} 
  = \frac{1}{2}\Bigl(p(\mu)^2 - (\rd_1\rd_sF)p(\mu) 
      - f + \sqrt{D}\Bigr), 
\label{dDKP-constraint-sol}
\eeq
where 
\beqnn
  p(\mu) = \rd_1S(\mu), \quad 
  D = \Bigl(p(\mu)^2 - (\rd_1\rd_sF)p(\mu) - f\Bigr)^2 
      - 4e^{\rd_s^2F}, 
\eeqnn
and $f$ denotes the right hand side 
of (\ref{dDKP-Hirota(iii)-S}).  
This enables one to eliminate $e^{\rd_sS(\lambda)}$ 
and $e^{\rd_sS(\mu)}$ from (\ref{dDKP-Hirota(i)-S}).  
The outcome is an equation for $D(\lambda)D(\mu)F$, 
$\rd_1S(\lambda)$ and $\rd_1S(\mu)$.  
One can rewrite it in a form 
that amounts to (\ref{dKP-HJ-gen}).  
Thus $S(\mu)$ anyhow turns out 
to satisfy a set of 
Hamilton-Jacobi equations of the form 
\beq
  \rd_nS(\mu) = H_n(\rd_1S(\mu)).  
\label{dDKP-HJ}
\eeq

A serious problem shows up here.  Namely, 
unlike the cases of the KP, Toda and 
BKP hierarchies, the right hand side 
of these Hamilton-Jacobi equations is 
not a polynomial function of $\rd_1S(\mu)$ 
but a fairly complicated (though algebraic) 
irrational function thereof.  Roughly speaking, 
$H_n(p)$ is a combination of polynomials and 
the square root of a quartic polynomial, 
namely,  
\beq
  H_n(p) = P_n(p) + Q_n(p)\sqrt{D}, 
\label{dDKP-Hamiltonian}
\eeq
where $P_n(p)$ and $Q_n(p)$ are polynomials 
in $p$, and $D$ is given by 
\beqnn
  D = (p^2 - (\rd_1\rd_sF)p - f)^2 - 4e^{\rd_s^2F}. 
\eeqnn
This new phenomena, hinting at the relevance 
of an elliptic curve, is closely related to 
the fact that the auxiliary linear problem 
(\ref{DKP-auxlin}) is essentially vector-valued. 
Because of this complicated structure of 
the Hamiltonians $H_n(p)$, the question 
of integrability of the underlying 
dispersionless system has not been resolved.  

These results could have been derived by 
the quasi-classical limit of the auxiliary 
linear equations.  Such an approach, however, 
is technically more subtle, because 
the auxiliary linear problem has matrix 
coefficients.  The foregoing approach 
based on the differential Fay identities 
is obviously simpler and more reliable.  
This is a demonstration of the advantage 
of using differential Fay identities.

\section{Concluding remarks}

One of the most remarkable results 
of our case studies is that differential 
(or difference) Fay identities are 
the auxiliary linear problem in disguise. 
This interpretation clarifies the meaning of 
the mysterious linear equations (\ref{KP-auxlin-SS}) 
of Sato and Sato \cite{SS82} and its generalization 
to other integrable hierarchies.  The status 
of differential Fay identities is thus parallel 
to that of dispersionless Hirota equations. 
We can use them as a new foundation of dispersive 
integrable hierarchies.  In this new framework, 
we might be able to go beyond dispersionless 
(or quasi-classical) limit to study higher orders 
of $\hbar$-expansion.  

Though the case studies in this paper 
are limited to the KP, Toda, BKP and DKP 
hierarchies, these results can be (and have been) 
generalized to some other cases.  
A particularly successful case \cite{TT07} 
is the charged multi-component KP hierarchy 
and its dispersionless analogue 
(which is identified to be the genus-zero 
universal Whitham hierarchy \cite{Krichever94}).  
In a sense, this is a multidimensional generalization 
of the Toda hierarchy as well.  The results 
on the KP and Toda hierarchies can be 
fully generalized to this case. 

Among the cases already studied, the DKP hierarchy 
and its dispersionless analogue 
are still posing many open problems.  
The technical complexity in this case stems from 
the fact that the auxiliary linear problem 
is essentially vector-valued.  In contrast, 
the charged multi-component KP hierarchy 
has actually a scalar-valued auxiliary linear problem, 
by which one can formulate its dispersionless analogue 
as the genus-zero universal Whitham hierarchy.  
Our preliminary consideration in this paper 
indicates that the dispersionless analogue 
of the DKP hierarchy might be related to 
the universal Whitham hierarchy of genus one 
\cite{Krichever94} or fall into one of 
examples of ``quasi-classical deformations 
of algebraic curves'' proposed by 
Kodama, Konopelchenko and Marninez Alonso 
\cite{KKMA04-05}. 
The Toda versions of the DKP hierarchy 
\cite{Willox05,GN05} will inherit 
a similar problem.  These issues deserve 
to be further studied. 

To conclude this paper, let us mention 
a geometric interpretation of dispersionless 
Hirota equations.  This interpretation 
is inspired by the recent work of 
Krichever, Marshakov and Zabrodin 
\cite{KMZ05} (see also the work of 
Carroll and Kodama \cite{CK95}).  
As the recent work of Eynard and Orantin 
\cite{EO07} on random matrices indicates, 
such a geometric point of view 
will be indispensable when one pursues 
higher orders of $\hbar$-expansion of 
$\log\tau_\hbar$.  
For simplicity, we now consider the case 
of the dispersionless KP hierarchy, but 
it is rather straightforward to generalize 
this interpretation to the dispersionless 
Toda hierarchy and the universal 
Whitham 
hierarchy of genus zero.  
A clue is to take the limit as $\mu \to \lambda$ 
in the dispersionless Hirota equation 
\beqnn
  e^{D(\lambda)D(\mu)F} 
  = \frac{p(\lambda) - p(\mu)}{\lambda - \mu}. 
\eeqnn
This yields the equation 
\beqnn
  e^{D(\lambda)^2F} = p'(\lambda), 
\eeqnn
by which one can rewrite 
the dispersionless Hirota equation as 
\beq
  \exp\Bigl(-\frac{1}{2}(D(\lambda)-D(\mu))^2F\Bigr) 
  = \frac{p(\lambda)-p(\mu)}
    {\sqrt{p'(\lambda)}\sqrt{p'(\mu)}(\lambda-\mu)}. 
\label{dKP-Hirota-pf}
\eeq
Remarkably, the quantity 
\beqnn
  E(\lambda,\mu) 
  = \frac{p(\lambda)-p(\mu)}
    {\sqrt{p'(\lambda)}\sqrt{p'(\mu)}}
\eeqnn
showing up on the right hand side 
is nothing but a local expression of 
the prime form \cite{Fay73,FK92} 
on the Riemann sphere in a neighborhood of 
$z = \infty$ with local coordinate $p(z)^{-1}$.  
We can thereby rewrite (\ref{dKP-Hirota-pf}) as 
\beq
  E(\lambda,\mu) 
  = (\lambda - \mu)
    \exp\Bigl(-\frac{1}{2}(D(\lambda)-D(\mu))^2F\Bigr). 
\label{dKP-primeform}
\eeq
This coincides with a formula presented 
by Krichever et al. \cite{KMZ05} 
for the case of nonzero genera.  
Moreover, taking the logarithmic 
second derivative of both hand sides 
with respect to $\lambda$ and $\mu$, 
we obtain an expression of 
the Bergmann kernel $B(\lambda,\mu) 
= \rd_\lambda\rd_\mu\log E(\lambda,\mu)$  
\cite{Fay73,FK92}: 
\beq
  B(\lambda,\mu) 
  = \frac{1}{(\lambda-\mu)^2} + D'(\lambda)D'(\mu)F. 
\label{dKP-Bergmann}
\eeq
Remarkably, this formula resembles 
a well known result on the two-point 
loop correlation function (which corresponds 
to $D'(\lambda)D'(\mu)F$ in this formula) 
of the Hermitian random matrix model 
in the large-$N$ limit.   This seems to 
indicates, as Eynard and Orantin \cite{EO07} 
observed in the case of random matrices, 
that the Bergmann kernel is one of fundamental 
building blocks of $\hbar$-expansion of 
$\log\tau_\hbar$.

\end{document}